\documentclass[extra,mreferee]{gji}
\usepackage{timet}

\usepackage{mathtools}
\usepackage{amssymb}
\usepackage{amsfonts}
\usepackage{amsmath}

\usepackage{graphicx}
\usepackage{epsf}
\usepackage{float}

\usepackage{booktabs}
\usepackage{color}
\usepackage{subfigure}
\usepackage{multicol}
\usepackage{enumerate}

\newcommand{\nb}{\nabla}
\newcommand{\bx}{{\boldsymbol{x}}}
\newcommand{\by}{{\boldsymbol{y}}}
\newcommand{\bz}{{\boldsymbol{z}}}
\newcommand{\bu}{{\boldsymbol{u}}}

\newcommand{\grdx}{\nabla_{\boldsymbol{x}}}

\newcommand{\e}{\mathrm{e}}

\newcommand{\tr}{\mathrm{Tr}}

\newcommand{\p}{\partial}
\newcommand{\f}[2]{\frac{#1}{#2}}

\newcommand{\sg}{\sigma}

\newcommand{\be}{\begin{equation}}
\newcommand{\ee}{\end{equation}}
\newcommand{\ba}{\begin{array}}
\newcommand{\ea}{\end{array}}
\newcommand{\bea}{\begin{eqnarray}}
\newcommand{\eea}{\end{eqnarray}}
\newcommand{\beas}{\begin{eqnarray*}}
\newcommand{\eeas}{\end{eqnarray*}}
\newcommand{\bl}{\begin{law}}
\newcommand{\el}{\end{law}}
\newcommand{\bthm}{\begin{thm}}
\newcommand{\ethm}{\end{thm}}

\newcommand{\ud}{\,\mathrm{d}}

\newcommand{\TT}{\mathrm{T}}

\newcommand{\bd}[1]{\boldsymbol{#1}}

\newcommand{\abs}[1]{\lvert#1\rvert}
\newcommand{\norm}[1]{\lVert#1\rVert}

\newcommand{\I}{\mathrm{i}}

\newcommand{\bP}{\boldsymbol{P}}
\newcommand{\bQ}{\boldsymbol{Q}}

\newcommand{\bp}{\boldsymbol{p}}
\newcommand{\bq}{\boldsymbol{q}}

\newcommand{\bF}{\mathbf F}

\newcommand{\cL}{\mathcal{L}}

\newcommand{\cO}{\mathcal{O}}

\newcommand{\barint}{\kern4pt \raise3.4pt\hbox{\vrule height.6pt
    width7pt} \kern-11pt \int}

\newcommand{\tp}{\mathrm{p}}
\newcommand{\ts}{\mathrm{s}}
\newcommand{\tsh}{\mathrm{sh}}
\newcommand{\tsv}{\mathrm{sv}}
\newcommand{\FGA}{\mathrm{F}}

\newcommand{\TR}{\mathrm{tr}}
\newcommand{\RE}{\mathrm{re}}
\newcommand{\IN}{\mathrm{in}}
\newcommand{\Id}{\mathrm{Id}}

\DeclareMathOperator{\Curl}{\nabla\times}

\newcommand{\ds}{\displaystyle}

\title[FGA for Elastic Wave Equation ]
  {Frozen Gaussian Approximation for 3-D Elastic Wave Equation and Seismic Tomography}
\author[J. C. Hateley, L. Chai, P. Tong, X. Yang]
{J. C. Hateley$^1$, L. Chai$^2$, P. Tong$^3$, X. Yang$^1$ \\
  $^1$ Department of Mathematics, University of California, Santa Barbara, CA 93106, USA\\
  $^2$  School of Mathematics, Sun Yat-sen University, Guangzhou, 510275, China \\
	$^3$ School of Physical and Mathematical Sciences \& Asian School of the Environment, \\ Nanyang Technological University, 637371, Singapore
}

\date{}
\pagerange{}
\volume{}
\pubyear{submitted}

\begin{document}

\maketitle

\begin{summary}
The purpose of this work is to generalize the frozen Gaussian approximation (FGA) theory to solve the 3-D elastic wave equation and use it as the forward modeling tool for seismic tomography with high-frequency data. FGA has been previously developed and verified as an efficient solver for high-frequency acoustic wave propagation (P-wave). The main contribution of this paper consists of three aspects: 1. We derive the FGA formulation for the 3-D elastic wave equation. Rather than standard ray-based methods ({e.g.} geometric optics and Gaussian beam method), the derivation requires to do asymptotic expansion in the week sense (integral form) so that one is able to perform integration by parts. Compared to the FGA theory for acoustic wave equation, the calculations in the derivation are much more technically involved due to the existence of both P- and S-waves, and the coupling of the polarized directions for SH- and SV-waves. In particular, we obtain the diabatic coupling terms for SH- and SV-waves, with the form closely connecting to the concept of Berry phase that is intensively studied in quantum mechanics and topology (Chern number). The accuracy and parallelizability of the FGA algorithm is illustrated by comparing to the spectral element method for 3-D elastic wave equation in homogeneous media; 2. We derive the interface conditions of FGA for 3-D elastic wave equation based on an Eulerian formulation and the Snell's law. We verify these conditions by simulating high-frequency elastic wave propagation in a 1-D layered Earth model. In this example, we also show that it is natural to apply the FGA algorithm to geometries with non-Cartesian coordinates; 3. We apply the developed FGA algorithm for 3-D seismic { wave-equation-based traveltime tomography and full waveform inversion, respectively}.
\end{summary}

\section{Introduction}
Images computed by seismic tomography can provide crucial information for the subsurface structures of Earth at different scales, and the understanding of tectonics, volcanism, and geodynamics~\citep[e.g.][]{Aki1976,Romanowicz1991,Rawlinson2010,Zhao2012a}. Wave-equation-based seismic tomography solves the nonlinear optimization problem iteratively for velocity models by computing seismograms and sensitivity kernels in 3-D complex models~\citep{Tromp2005,Liu2008,Liu2012,Tong2014a}. This leads to successful applications including imaging the velocity models of the southern California crust \citep{Tape2009,Tape2010}, the European upper mantle \citep{Zhu2012}, {the North Atlantic region \citep{Rickers2013}, and the Japan islands \citep{Simute2016}}. The performance of seismic tomography is restricted by how accurate one can solve the 3-D wave equation for synthetic seismograms and sensitivity kernels. The dominant frequency of a low-frequency earthquake is around 1 Hz, while the one of a typical earthquake is around 5 Hz \citep[e.g.][]{nakamichi2003source}, which leads to demanding and even unaffordable computational cost. Real applications usually chose significantly low-frequency data in order to be computationally feasible {\citep[e.g.][]{Tape2009,Zhu2012,Simute2016}}, since simulating high-frequency seismic waves requires much more powerful computational resources on both memories and CPU time than the low-frequency waves. This imposes a demand on improving the efficiency and accuracy of numerical methods for computing high-frequency waves in order to make use of real seismic data around their dominant frequencies.

In previous works \citep{chai2017frozen,chai2018tomo}, the authors have developed and verified frozen Gaussian approximation (FGA) as an efficient solver for computing high-frequency acoustic wave propagation (P-wave), {wave-equation-based traveltime tomography} and full waveform inversion (FWI). The core idea of FGA is to approximate seismic wavefields by fixed-width Gaussian wave packets, whose dynamics follow ray paths with the prefactor amplitude equation derived delicately from an asymptotic expansion on phase plane. With a multicore processors computer station, the property that the FGA algorithm is embarrassingly parallel makes possible the application of FGA to compute 3-D {high frequency sensitivity kernels, and further used for 3-D traveltime tomography} and FWI. Compared to other ray-based methods including WKBJ \citep[e.g.][]{Chapman:76,ChDr:82}, WKM \citep[e.g.][]{NiDiHe:00,HeNi:05}, generalized ray theory \citep[e.g.][]{He:68,ViHe:88}, {seismic traveltime tomography} \citep[e.g.][]{Aki1977, tong2017time}, Kirchhoff migration \cite[e.g.][]{Gr:86,KeBe:88} and Gaussian beam
migration \citep[e.g.][]{Hi:90,Hi:01,SEG-2003-11141117,Gr:05,GrBe:09,PoSePoVe:10}, FGA does not need to solve ray paths by shooting to reach the receivers, and provides accurate solutions at the presence of caustics and multipathing, with no requirement on tuning beam width parameters to achieve a good resolution \citep[]{CePoPs:82,Hi:90,FoTa:09,QiYi:10,LuYa:11}.

In this paper, we first generalize the FGA theory to solve the 3-D elastic wave propagation. Different from the WKBJ theory or Gaussian beam method which can be derived by direct asymptotic expansion, the derivation of FGA formulation requires to do the asymptotic expansion in an integral form (weak sense) so that one is able to perform integration by parts to eliminate the extra constraints yielded by direct asymptotic expansion. Compared to the previous works on FGA \cite[]{LuYa:CPAM,chai2017frozen,chai2018tomo}, the calculations for the derivation are much more technically involved due to the existence of both P- and S-waves. Particularly, the diabatic coupling of the polarized directions for SH- and SV-waves leads to a term closely connected to the concept of Berry phase which is intensively studied in quantum mechanics and topology (Chern number) \cite[e.g.][]{Be:84,Si:83}. We prove the accuracy and parallelizability of the FGA algorithm by comparing to the spectral element method \citep[e.g.][]{Komatitsch1999,Komatitsch2005,Tromp2008} for the 3-D elastic wave equation in homogeneous media, where one has the analytical solution as a benchmark. We derive the interface conditions of FGA for the 3-D elastic wave equation based on an Eulerian formulation and the Snell's law. These interface conditions are verified by simulating high-frequency elastic wave propagation in a 1-D layered Earth model. We also show how natural to apply the FGA algorithm to geometries with non-Cartesian coordinates in the simulation of this model.

The second part of this paper contributes to the study of seismic tomography with high-frequency data. {Wave-equation-based traveltime tomography} and FWI are able to generate higher-resolution images of the {Earth's} interior than the conventional ray-based tomography method \citep[][]{Virieux2009}. Especially, some recent novel studies on FWI using optimal transport distance (e.g. Wasserstein metric) are able to capture traveltime differences between seismic signals, and thus overcomes the cycling effect in standard FWI method \citep[e.g.][]{EnFr:14,MeBrMeOuVi:16,YaEnSuHa:18,ChChWuYa:arXiv}. However, seismic tomography using high-frequency data increases computational cost drastically, which restricts the application of these wave-equation-based inversion methods, and only low-frequency data are modeled and inverted in some real applications \citep[e.g.][]{Tape2007, Zhu2012, Simute2016}. To address the computation challenge, we use the developed FGA algorithm to compute the 3-D high-frequency sensitivity kernels for { wave-equation-based traveltime tomography and FWI}, respectively. We apply FGA to both {wave-equation-based traveltime tomography} and FWI on synthetic crosswell seismic data with dominant frequencies as high as those of real crosswell data, and use a hierarchical approach which first uses {traveltime tomography} to create a macro-scale model and then adopts FWI to generate a high-resolution micro-scale model.


\section{Frozen Gaussian approximation}
We derive the frozen Gaussian approximation (FGA) formulation for the following inhomogeneous, isotropic elastic wave equation:
\begin{equation}\label{eq:ewe}
\rho(\bx)\partial^2_t\bu = \bigl(\lambda(\bx) + \mu(\bx)\bigr)\nabla(\nabla\cdot \bu ) + \mu(\bx)\Delta\bu + \bF(t,\bx),\quad \bx\in\mathbb{R}^3,
\end{equation}
where $\bu(t,\bx)$ is the wavefield, $\rho(\bx)$ is the density, $\lambda(\bx)$ and $\mu(\bx)$ are the Lam\'{e} parameters, and $\bF(t,\bx)$ is the body force. We shall assume $\bF=0$ in this section as a sake of simplicity for presenting the FGA formula. Remark that eq.~\eqref{eq:ewe} can be rewritten as,
\begin{equation}\label{eq:ewe_ps}
\partial^2_t\bu = c_{\tp}^2\nabla(\nabla\cdot \bu) - c_{\ts}^2\nabla\times\nabla \times \bu +\frac{\bd{F}}{\rho},
\end{equation}
with the P-wave, and S-wave speeds defined as,
\begin{equation}\label{eq:ps_speed}
c^2_{\tp}(\bx) = \frac{\lambda(\bx) + 2\mu(\bx)}{\rho(\bx)}\indent c^2_{\ts}(\bx) = \frac{\mu(\bx)}{\rho(\bx)}.
\end{equation}
We consider the elastic wave eq.~\eqref{eq:ewe}, with the following initial conditions
\begin{equation}\label{eq:ini}
\begin{cases}
~\bu(0,\bx)   = \bd{f}^k(\bx), \\
~\partial_t\bu(0,\bx) = \bd{g}^k(\bx). \\
\end{cases}
\end{equation}
Note that all the bolded variables and parameters will be referred to vectors in $\mathbb{R}^3$ without any further clarification.


 \subsection{Solution Ansatz} The FGA approximates the wavefield $\bu(t,\bx)$ in eq.~\eqref{eq:ewe} by a summation of dynamic frozen Gaussian wave packets,
  \begin{equation}\label{eq:FGA}
  \begin{aligned}
     u^{k}_{\FGA}(t, \bd{x}) & \approx \sum_{(\bd{q},\bd{p})\in G_\pm^\tp} \frac{a_\tp\bd{\hat{N}}_\tp
      \psi^k_\tp}{(2\pi/k)^{9/2}}
    e^{{\I}{k}\bd{P}_\tp\cdot(\bd{x} - \bd{Q}_\tp) -
      \frac{k}{2} \abs{\bd{x}
        - \bd{Q}_\tp}^2}\delta\bd{q}\delta \bd{p}  \\
    & + \sum_{(\bd{q},\bd{p})\in G_\pm^\ts} \frac{a_\ts\bd{\hat{N}}_\ts\psi^k_\ts}{(2\pi /k)^{9/2}}
    e^{{\I}{k} \bd{P}_\ts\cdot(\bd{x} - \bd{Q}_\ts) -
      \frac{k}{2} \abs{\bd{x} - \bd{Q}_\ts}^2}\delta\bd{q}\delta \bd{p},
  \end{aligned}
\end{equation}
with the weight functions
\begin{align}\label{eq:psi}
&  \psi^k_{\tp,\ts}(\bd{q}, \bd{p}) =
   \int \alpha^k_{\tp,\ts}(\bd{y},\bd{q},\bd{p})
  e^{- \I k \bd{p}\cdot(\bd{y} - \bd{q}) -
    \frac{k}{2} \abs{\bd{y} - \bd{q}}^2} \ud\bd{y},\\ \label{eq:alpha^k}
&\alpha^k_{\tp,\ts}(\bd{y},\bd{q},\bd{p})=\frac{1}{2kc_{\tp,\ts}\abs{\bp}^3}\Bigl(k\bd{f}^k(\bd{y})c_{\tp,\ts}\abs{\bd{p}}\pm
  \I\bd{g}^k(\bd{y}) \Bigr)\cdot\bd{\hat{n}}_{\tp,\ts}.
\end{align}
In eq.~\eqref{eq:FGA}, $\I=\sqrt{-1}$ is the imaginary unit,
and we use unbolded subscripts/superscripts ``$\tp$'' and ``$\ts$'' to indicate P- and S-waves respectively, and superscripts $k$ to indicate quantities depending on wave number $k$. The quantities that have both $\tp$ and $\ts$ as subscripts/superscripts mean that they can be referred to both P- and S-waves. Here $G_\pm^{\tp,\ts}$ refers to the initial sets of Gaussian center $\bd{q}$ and propagation vector $\bd{p}$ for P- and S-waves respectively, and $\pm$ indicates the two-way wave propagation directions correspondingly. $\bd{\hat{N}}_{\tp,\ts}(t)$ are unit vectors indicating the polarized directions of P- and S-waves, i.e. if $(\bd{q},\bd{p})\in G_\pm^{\tp}$, then $\bd{\hat{N}}_\tp \parallel \bd{P}_\tp$; and if $(\bd{q},\bd{p})\in G_\pm^{\ts}$, then $\bd{\hat{N}}_\ts\perp\bd{P}_\ts$. In \eqref{eq:alpha^k}, $\bd{\hat{n}}_{\tp,\ts}$ are the initial directions of P- and S-waves, i.e. $\bd{\hat{n}}_{\tp,\ts}=\bd{\hat{N}}_{\tp,\ts}(0)$, and the ``$\pm$'' on the right-hand-side of the equation indicate that the $\alpha^k_{\tp,\ts}$ correspond to $(\bd{q},\bd{p})\in G_\pm^{\tp,\ts}$. {Remark that
$e^{\I k \bd{p}\cdot(\bd{y} - \bd{q}) -
    \frac{k}{2} \abs{\bd{y} - \bd{q}}^2}$ can be understood as complex localized wave-packet centered at $\bd{q}$ with propagation vector $\bd{p}$, and
   $\psi^k_{\tp,\ts}$ is the projection of the initial wavefield onto each wave-packet computed by eq.~\eqref{eq:psi} with $\bd{y}$ serving as the dummy variable in the integration.}
Associated with each frozen Gaussian wave-packet, the \emph{time-dependent} quantities are the position center $\bQ_{\tp,\ts}(t,\bd{q},\bd{p})$, momentum center $\bP_{\tp,\ts}(t,\bd{q},\bd{p})$, amplitude $a_{\tp,\ts}(t,\bd{q},\bd{p})$ and unit direction vectors $\bd{\hat{N}}_{\tp,\ts}(t)$. Note that all the S-waves discussed in the formulation will include both SH- and SV-waves.

\subsection{Formulation and Algorithm}
The derivation of the FGA formulation involves with the asymptotic expansion in the integral form (weak sense), with proper integration by parts performed to convert powers of distance to the Gaussian center $\bd{x}-\bd{Q}$ to orders of the wavelength $k^{-1}$. It is quite lengthy and technically involved, and thus we leave it to Appendix~\ref{app:FGA} for the readers who are interested in the mathematical details, and only present the FGA formulation as below.

For the ``+'' wave propagation, i.e., $(\bd{q},\bd{p})\in G_+^{\tp,\ts}$, the Gaussian center $\bd{Q}_{\tp,\ts}(t,\bd{q},\bd{p})$ and propagation vector $\bd{P}_{\tp,\ts}(t,\bd{q},\bd{p})$ follow the ray dynamics
\begin{equation}\label{eq:char_plus}
  \begin{cases}
    \displaystyle
    \frac{\ud \bd{Q}_{\tp,\ts}}{\ud t} = c_{\tp,\ts}(\bd{Q}_{\tp,\ts})
\frac{\bd{P}_{\tp,\ts}}{\abs{\bd{P}_{\tp,\ts}}},\\[.5em]
    \displaystyle
    \frac{\ud \bd{P}_{\tp,\ts}}{\ud t} =- \partial_{\bd{Q}} c_{\tp,\ts}(\bd{Q}_{\tp,\ts})
\abs{\bd{P}_{\tp,\ts}},
  \end{cases}
\end{equation}
with initial conditions
\begin{equation}\label{eq:ini_plus}
\bd{Q}_{\tp,\ts}(0, \bd{q}, \bd{p}) =
\bd{q} \quad \text{and} \quad
\bd{P}_{\tp,\ts}(0, \bd{q}, \bd{p}) = \bd{p}.
\end{equation}

For the ``-'' wave propagation, i.e., $(\bd{q},\bd{p})\in G_-^{\tp,\ts}$,  the Gaussian center $\bd{Q}_{\tp,\ts}(t,\bd{q},\bd{p})$ and propagation vector $\bd{P}_{\tp,\ts}(t,\bd{q},\bd{p})$ follow the ray dynamics
\begin{equation}\label{eq:char_minus}
  \begin{cases}
    \displaystyle
    \frac{\ud \bd{Q}_{\tp,\ts}}{\ud t} = -c_{\tp,\ts}(\bd{Q}_{\tp,\ts})
\frac{\bd{P}_{\tp,\ts}}{\abs{\bd{P}_{\tp,\ts}}},\\[.5em]
    \displaystyle
    \frac{\ud \bd{P}_{\tp,\ts}}{\ud t} = \partial_{\bd{Q}} c_{\tp,\ts}(\bd{Q}_{\tp,\ts})
\abs{\bd{P}_{\tp,\ts}},
  \end{cases}
\end{equation}
with initial conditions
\begin{equation}\label{eq:ini_minus}
\bd{Q}_{\tp,\ts}(0, \bd{q}, \bd{p}) =
\bd{q} \quad \text{and} \quad
\bd{P}_{\tp,\ts}(0, \bd{q}, \bd{p}) = \bd{p}.
\end{equation}
Remark that, the equations for ``-'' wave propagation \eqref{eq:char_minus} have opposite signs of the right-hand side to the equations for ``+'' wave propagation \eqref{eq:char_plus}, and other than that, they are the same. Actually they can be both viewed as the Hamiltonian system with $H(\bd{Q},\bd{P})=\pm c_{\tp,\ts}(\bd{Q})\abs{\bd{P}}$ as the Hamiltonian function for ``+'' and ``-'' wave propagation, respectively.

The prefactor amplitudes $\bd{a}_{\tp,\ts}(t,\bd{q},\bd{p})$ satisfy the following equations, where S-waves have been decomposed into SH- and SV-waves,
\begin{align}\label{eq:amp_P}
&\frac{\ud a_{\tp}}{\ud t} =  a_{\tp}\biggl(\pm\frac{\partial_{\bd{Q}_{\tp}}c_{\tp}\cdot \bd{P}_{\tp}}{|\bd{P}_{\tp}|} + \frac{1}{2}\tr\Bigl(Z_{\tp}^{-1}\frac{\ud Z_{\tp}}{\ud t}\Bigr)\biggr), \\
& \frac{\ud a_{\tsv}}{\ud t} =  a_{\tsv}\biggl(\pm\frac{\partial_{\bd{Q}_{\ts}}c_{\ts}\cdot \bd{P}_{\ts}}{|\bd{P}_{\ts}|} + \frac{1}{2}\tr\Bigl(Z_{\ts}^{-1}\frac{\ud Z_{\ts}}{\ud t}\Bigr)\biggr)-a_{\tsh}\frac{\ud \bd{\hat{N}}_{\tsh}}{\ud t}\cdot\bd{\hat{N}}_\tsv, \label{eq:amp_SV}\\
& \frac{\ud a_{\tsh}}{\ud t} = a_{\tsh}\biggl(\pm\frac{\partial_{\bd{Q}_{\ts}}c_{\ts}\cdot \bd{P}_{\ts}}{|\bd{P}_{\ts}|} + \frac{1}{2}\tr\Bigl(Z_{\ts}^{-1}\frac{\ud Z_{\ts}}{\ud t}\Bigr)\biggr)+a_{\tsv}\frac{\ud \bd{\hat{N}}_{\tsh}}{\ud t}\cdot\bd{\hat{N}}_\tsv, \label{eq:amp_SH}
\end{align}
with the initial conditions $a_{\tp,\tsv,\tsh}=2^{3/2}$, and $\bd{\hat{N}}_{\tsv}$ and $\bd{\hat{N}}_{\tsh}$ are the two unit directions perpendicular to $\bd{P}_\ts$, referring to the polarized directions of SV- and SH-waves, respectively. Here ``$\pm$'' corresponds to the two-way wave propagation directions, and we have used the short-hand notations,
 \begin{equation}\label{eq:op_zZ_app}
  \partial_{\bd{z}}=\partial_{\bd{q}}-\I\partial_{\bd{p}},
  \qquad
  Z_{\tp,\ts}=\partial_{\bd{z}}(\bd{Q}_{\tp,\ts}+\I\bd{P}_{\tp,\ts
 }).
\end{equation}
Note that, the prefactor equation \eqref{eq:amp_P} is consistent with the one for acoustic wave equation with $c^2=(\lambda+2\mu)/\rho$ \citep{chai2017frozen} , and the last terms on the right-hand-side of \eqref{eq:amp_SV}-\eqref{eq:amp_SH} indicate the diabatic coupling of the polarized directions for SH- and SV-waves, which are closely connected to the concept of Berry phase intensively studied in quantum mechanics and topology (Chern number) \cite[e.g.][]{Be:84,Si:83}.

\begin{figure}
\centering
\includegraphics*[scale = .5]{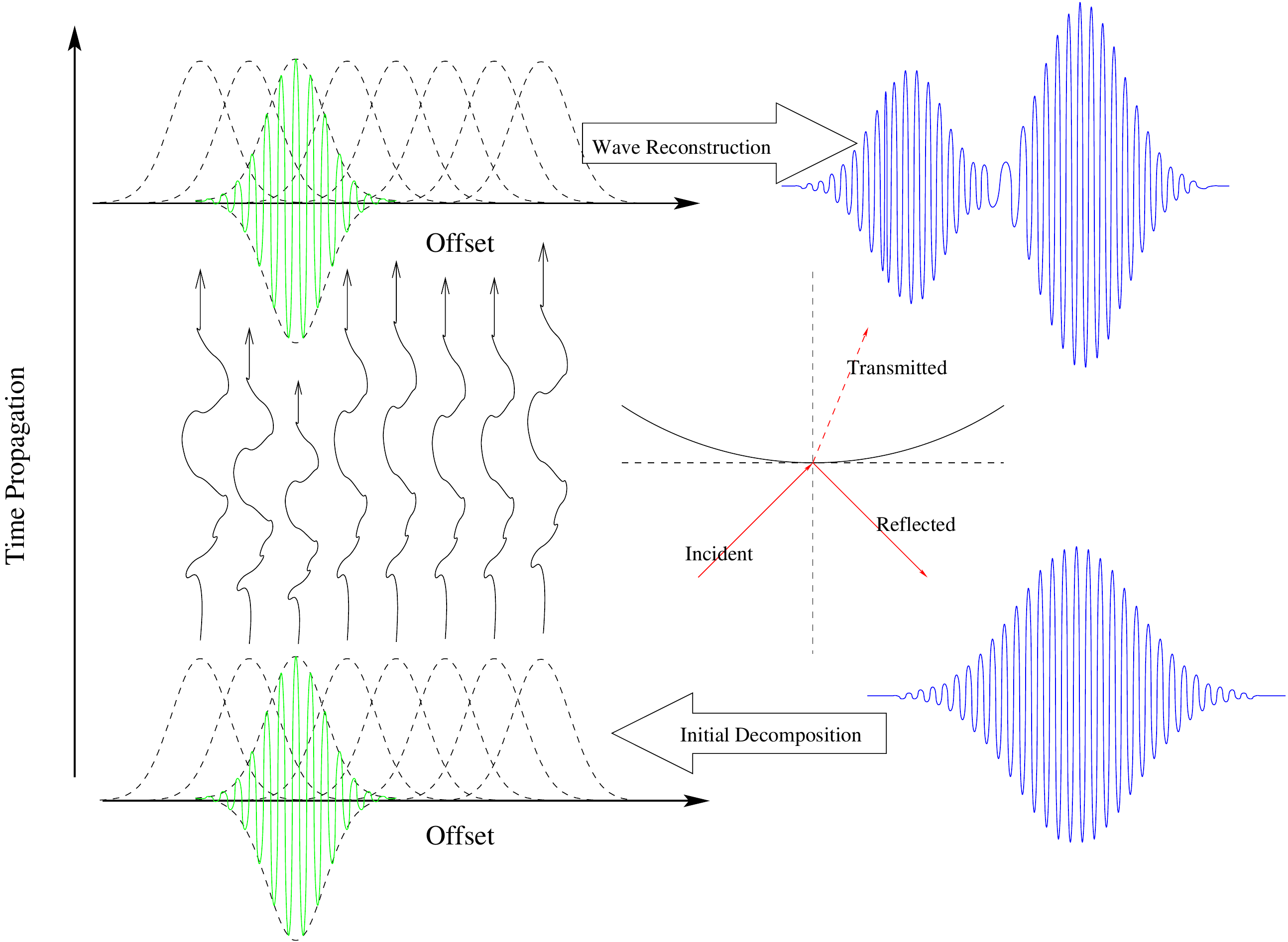}
\caption{This cartoon figure illustrates the main three steps of the FGA algorithm: Step 1, Initial decomposition, i.e., to choose the proper sets $G^{\tp,\ts}_{\pm}$ by the fast FBI transform algorithm \citep[Page 2]{YaLuFo:13} to decompose the initial wavefields into Gaussian wave packets, and calculate the corresponding weight function $\psi^k_{\tp,\ts}$ defined in eq.~\eqref{eq:psi}; Step 2, Time propagation, i.e., to solve by some numerical solver eqs.~\eqref{eq:char_plus}, \eqref{eq:char_minus}, \eqref{eq:amp_P}, \eqref{eq:amp_SV} and  \eqref{eq:amp_SH} for the dynamics of Gassian center, propagation vector and prefactor amplitude. Transmission and reflection conditions are needed at the presence of interfaces as given in Section~\ref{sec:TR_cond}; Step 3, Wave reconstruction, i.e., to compute eq.~\eqref{eq:FGA} for the elastic wavefield at time $T$. The fast Gaussian summation algorithm proposed in \citet[Section 3.3]{chai2018tomo} is highly recommended if a large number of reconstruction grid points are needed.  }\label{fig:FGA_alg}
\end{figure}

\noindent{\it Algorithm.} A flowchart is shown in Fig.~\ref{fig:FGA_alg} to describe the FGA algorithm, with the technique details discussed in the figure caption.


\subsection{Accuracy and Parallelizability}
We check the accuracy and parallelizability of the FGA method by simulating the elastic wave propagation at a homogeneous media with absorbing boundary conditions, for which one can solve the analytical solution as a benchmark. Specifically, we consider
\begin{equation}\label{eq:ex_homo}
\rho\partial^2_t\bd{u} = (\lambda + \mu)\nabla(\nabla\cdot \bd{u} ) + \mu\Delta\bd{u} + \bd{F}(t)\delta(\bd{x}-\bd{x}_0),
\end{equation}
with $\bd{x}_0 = (x_{1,0},x_{2,0},x_{3,0})$ as the source location, $\bd{F} = \bigl(F_1(t), F_2(t), F_3(t)\bigr)^\TT$ as the source time function at $\bd{x}_0$, and $\rho$, $\lambda$ and $\mu$ as constants. Eq.~\eqref{eq:ex_homo} has the following analytical solution
\begin{equation}\label{eq:ex_anal}
\begin{aligned}
u_i = \ds\frac{(x_i-x_{i,0})(x_j-x_{j,0})}{4\pi\rho c_{\tp}^2r^3}F_j(t - r/c_{\tp}) + \frac{r^2\delta_{ij} - (x_i-x_{i,0})(x_j-x_{j,0})}{4\pi\rho c_{\ts}^2r^3}F_j(t - r/c_{\ts})  \\
+ \frac{3(x_i-x_{i,0})(x_j-x_{j,0}) - r^2\delta_{ij}}{4\pi\rho r^5}\int_{r/c_{\tp}}^{r/c_{\ts}}sF_j(t - s)\ \ud s,
\end{aligned}
\end{equation}
where we have used the Einstein's index summation convention, $\bd{u}=\bigl(u_1,u_2,u_3\bigr)^\TT$, $r = \|\bx - \bx_0\|_2$ and $\delta_{ij}$ is the Kronecker delta function. Here we use $\bd{x}=(x_1,x_2,x_3)^\TT$ for a simple formulation of the analytical solution, which is actually $\bd{x}=(x,y,z)^\TT$ in standard notations.

In the numerical tests, we simulate the elastic wave propagation in a domain of the size $128$ km $\times$ $128$ km $\times$ $128$ km, and take the source time function as
\begin{equation*}
F_j(t) = \cos\left(2\pi f(t+T_0)\right)\exp\left(-(t+T_0)^2/\sigma^2 \right),
\end{equation*}
with $f$ as the frequency, $T_0=0.1768$ s and $\sigma=0.8660$ s. We choose P- and S-wave speeds as $c_{\tp} = 8$ km/s, $c_{\ts} = 4.619$ km/s, density $\rho=1\;\text{kg/km}^3$, final propagation time $T = 13.86$ s and the source location $\bd{x}_0 = (64, 64, 64)$ km. We use a fourth order Runge-Kutta (RK4) with fixed time step as the numerical solver for eqs.~\eqref{eq:char_plus}, \eqref{eq:char_minus}, \eqref{eq:amp_P}, \eqref{eq:amp_SV} and  \eqref{eq:amp_SH}. To compare the performance of the FGA method to {the} spectral element software package SPECFEM3D\footnote{https://geodynamics.org/cig/software/specfem3d/}, simluations are done on a cluster of each node equipped with an Intel(R) Xeon(R) E5-2670 (2.60GHz) and a total of 64GB RAM. The accuracy and parallelizability of FGA are tested for $f=1.4702$ Hz, and the comparison of computational time to SPECFEM3D is done for a range of $f$ from $0.3676$ Hz to $11.7617$ Hz.

\begin{figure}
\subfigure[Initial wavefield]
{\begin{minipage}[t]{0.33\linewidth}
\centering
\includegraphics*[scale = .25]{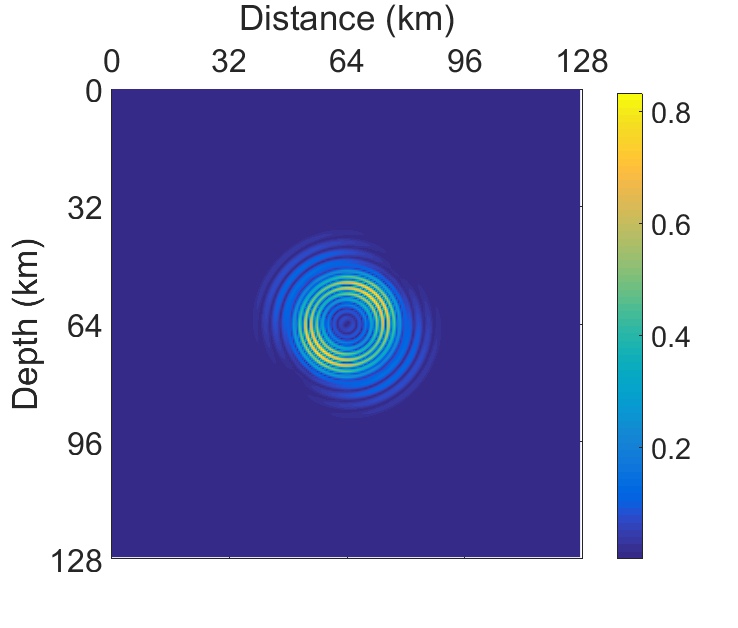}
\end{minipage}}
\subfigure[$t = 6.93$ s]
{\begin{minipage}[t]{0.33\linewidth}
\centering
\includegraphics*[scale = .25]{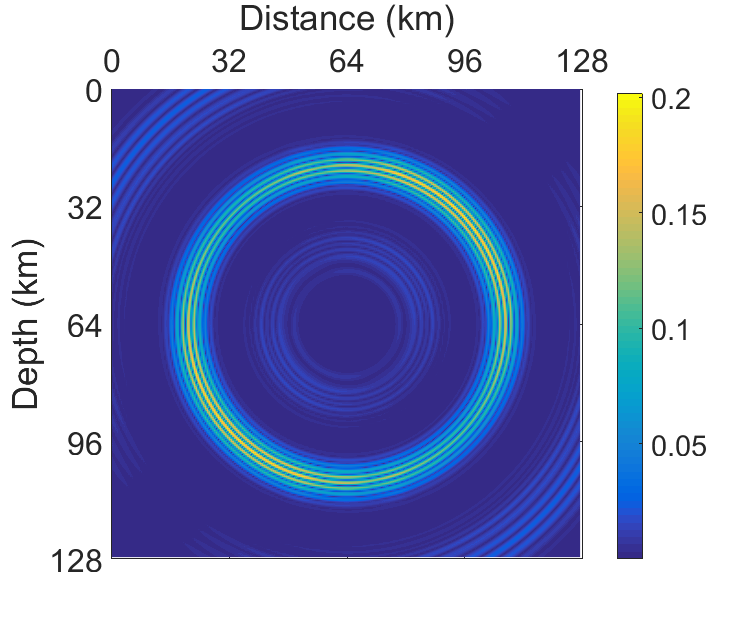}
\end{minipage}}
\subfigure[ $t = 13.86$ s]
{\begin{minipage}[t]{0.33\linewidth}
\centering
\includegraphics*[scale = .25]{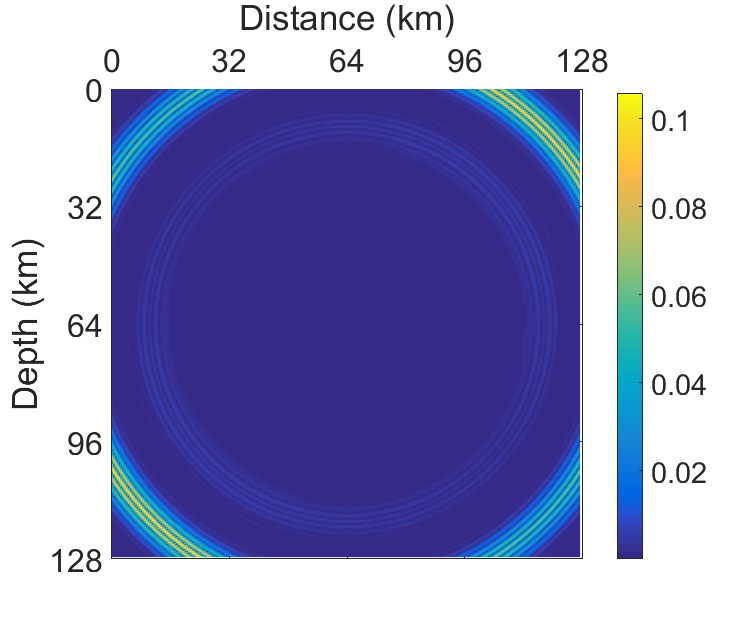}
\end{minipage}}
\caption{Modulus of the elastic wavefield computed by the FGA method in homogeneous media, with the analytical solution given by eq. \eqref{eq:ex_anal}. The subfigures from left to right show the slices of $\norm{\bd{u}}_2=\sqrt{u_1^2+u_2^2+u_3^2}$ at $y = 64$km for $t=0, 6.93, 13.86$ s. The frequency of the source time function is $f = 1.4702$ Hz. }\label{fig:ex1_dynamics}
\end{figure}

\begin{figure*}
\subfigure[$u_1$]
{\begin{minipage}[t]{0.33\linewidth}
\centering
\includegraphics*[scale = .24]{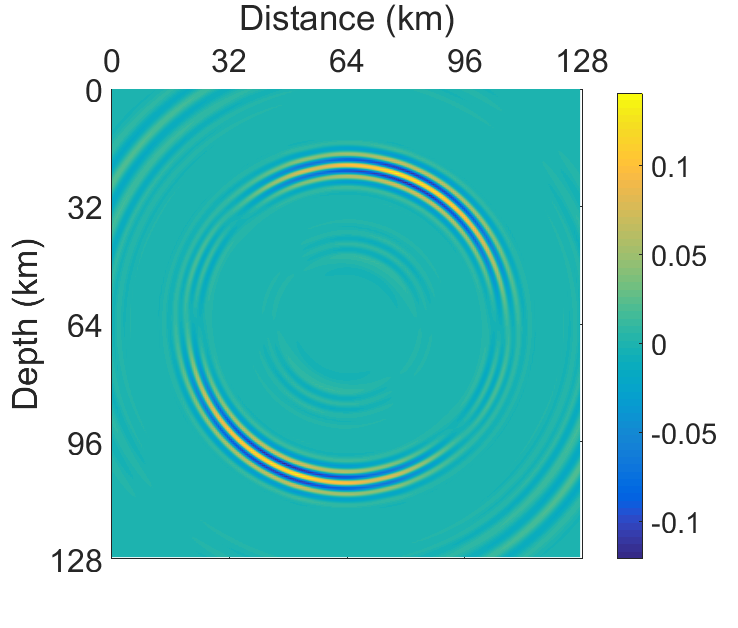}
\end{minipage}}
\subfigure[$u_1$: P-wave component]
{\begin{minipage}[t]{0.33\linewidth}
\centering
\includegraphics*[scale = .24]{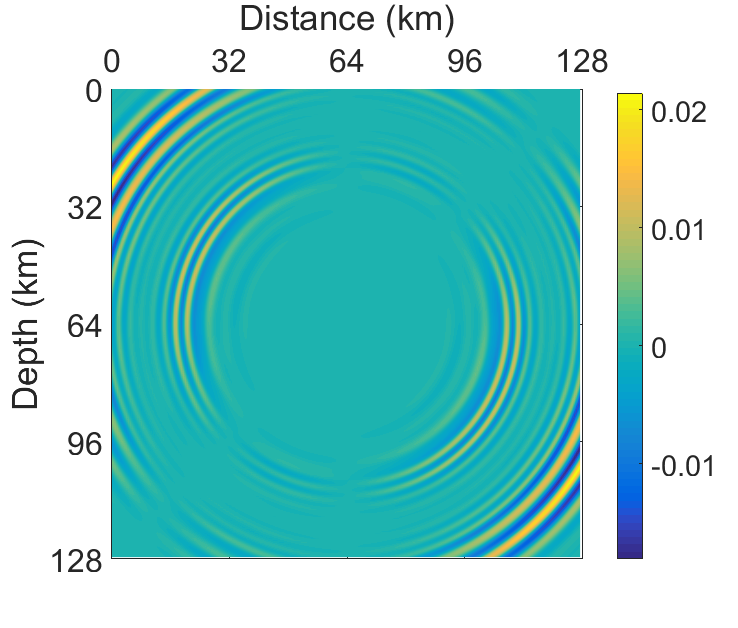}
\end{minipage}}
\subfigure[$u_1$: S-wave component]
{\begin{minipage}[t]{0.33\linewidth}
\centering
\includegraphics*[scale = .24]{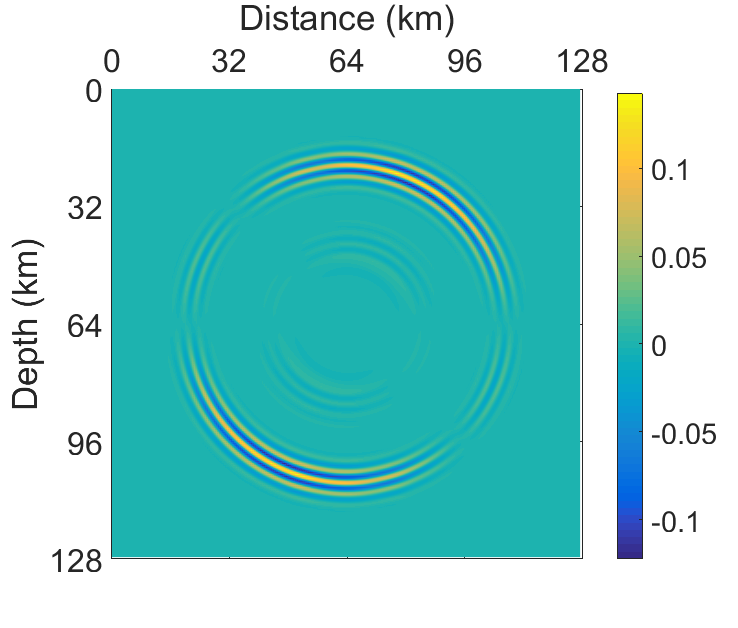}
\end{minipage}}\\
\subfigure[$u_2$]
{\begin{minipage}[t]{0.33\linewidth}
\centering
\includegraphics*[scale = .25]{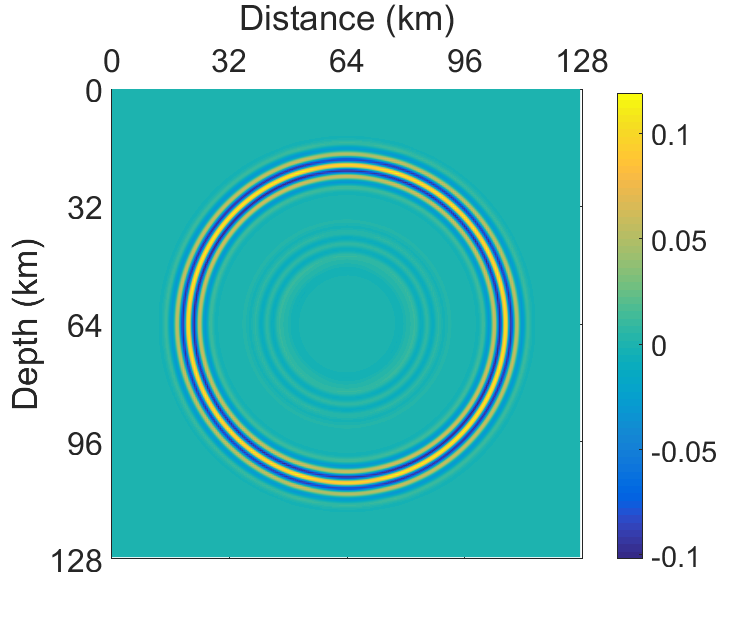}
\end{minipage}}
\subfigure[$u_2$: P-wave component]
{\begin{minipage}[t]{0.33\linewidth}
\centering
\includegraphics*[scale = .25]{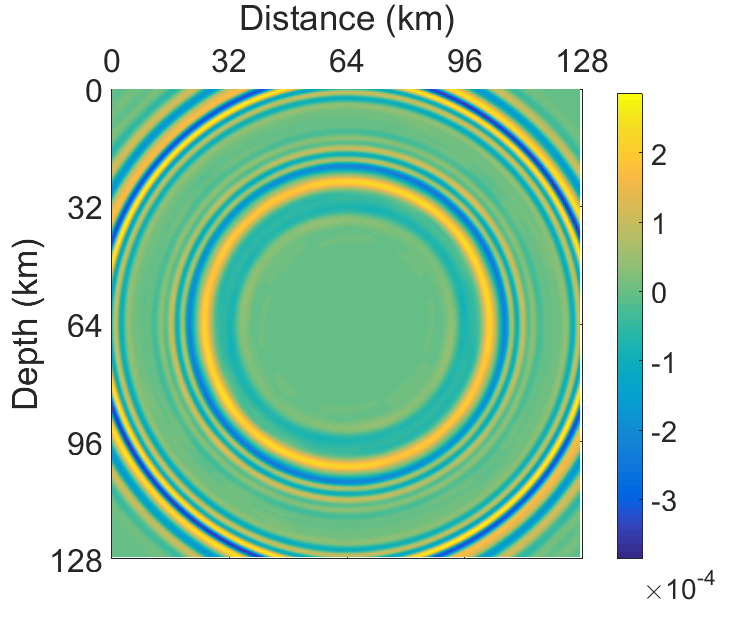}
\end{minipage}}
\subfigure[$u_2$: S-wave component]
{\begin{minipage}[t]{0.33\linewidth}
\centering
\includegraphics*[scale = .25]{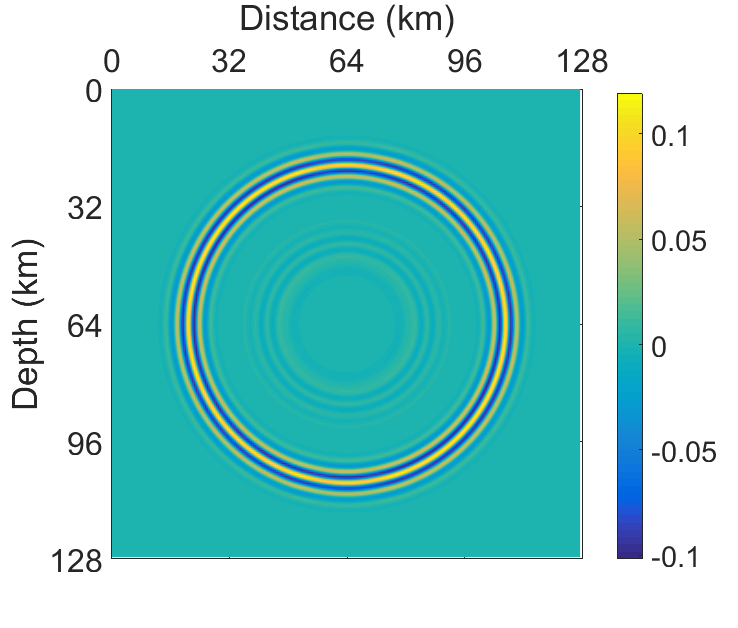}
\end{minipage}}\\
\subfigure[$u_3$]
{\begin{minipage}[t]{0.33\linewidth}
\centering
\includegraphics*[scale = .25]{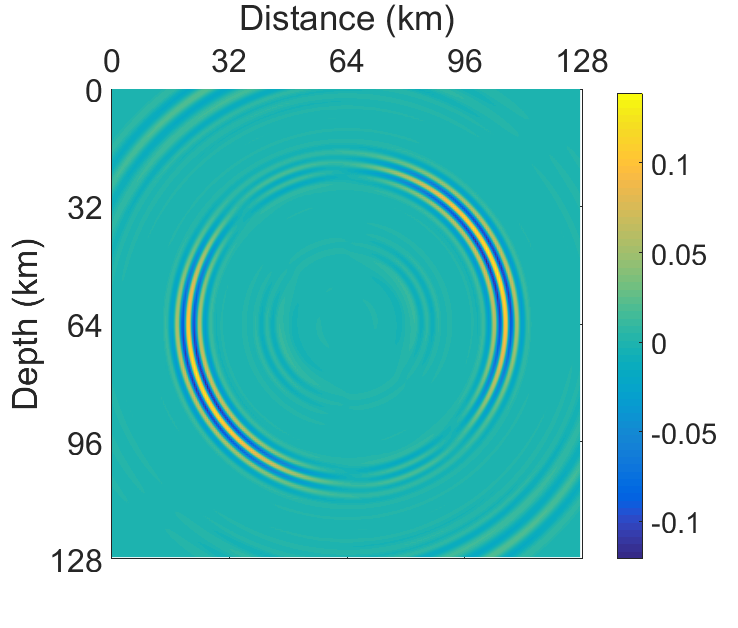}
\end{minipage}}
\subfigure[$u_3$: P-wave component]
{\begin{minipage}[t]{0.33\linewidth}
\centering
\includegraphics*[scale = .25]{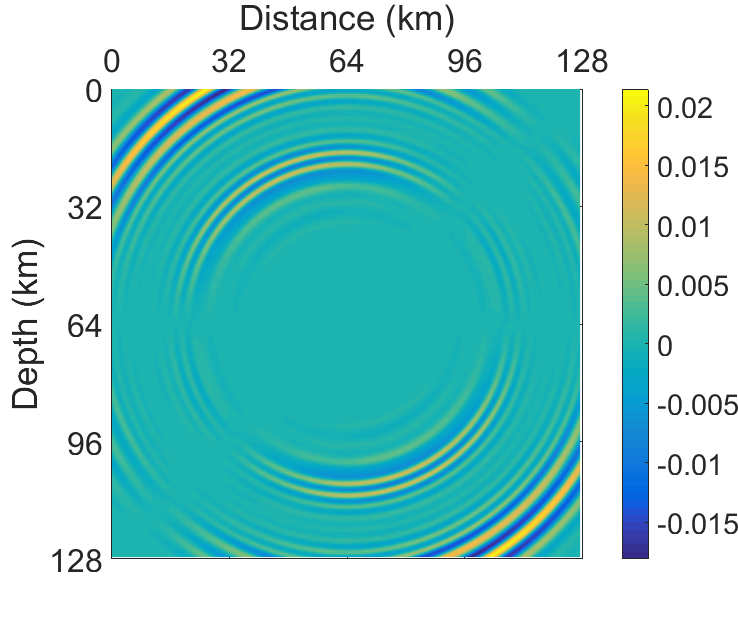}
\end{minipage}}
\subfigure[$u_3$: S-wave component]
{\begin{minipage}[t]{0.33\linewidth}
\centering
\includegraphics*[scale = .25]{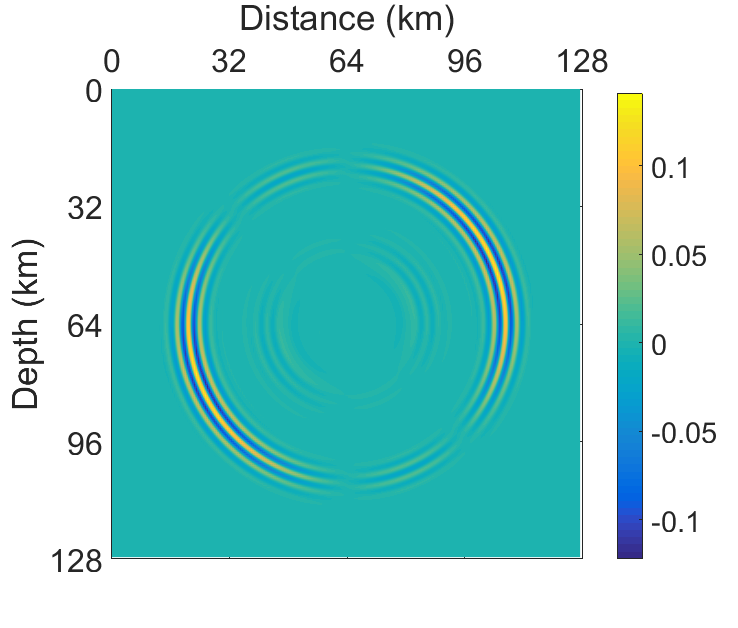}
\end{minipage}}\\
\caption{Modulus of the componentwise elastic wavefield computed by the FGA method in homogeneous media, with the analytical solution given by eq. \eqref{eq:ex_anal}. The subfigures show the components of $\bd{u}=(u_1,u_2,u_3)^\TT$ for $t = 6.93$ s and the source frequency $f = 1.4702$ Hz. The second and third columns show the computed P- and S-wavefields by the FGA method, respectively. }\label{fig:ex1_components}
\end{figure*}

\begin{figure*}
\subfigure[FGA]
{\begin{minipage}[t]{0.33\linewidth}
\centering
\includegraphics*[scale = .25]{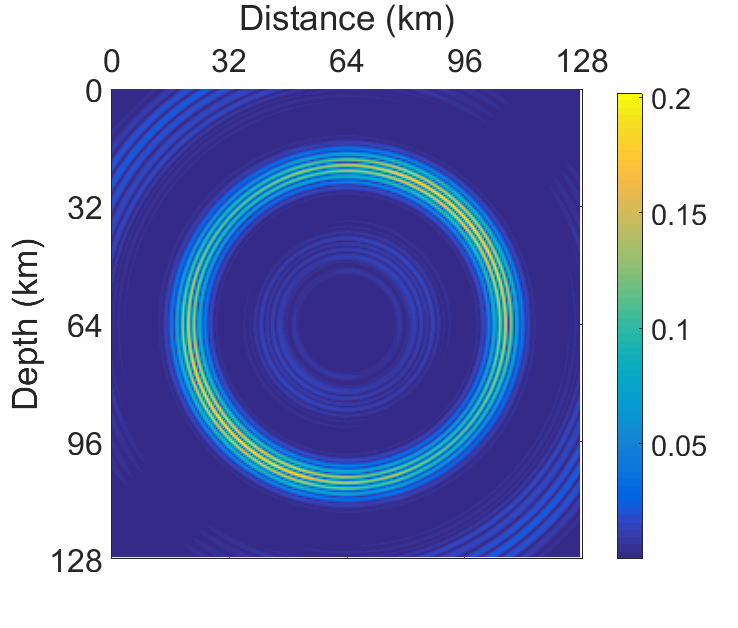}
\end{minipage}}
\subfigure[SEM]
{\begin{minipage}[t]{0.33\linewidth}
\centering
\includegraphics*[scale = .25]{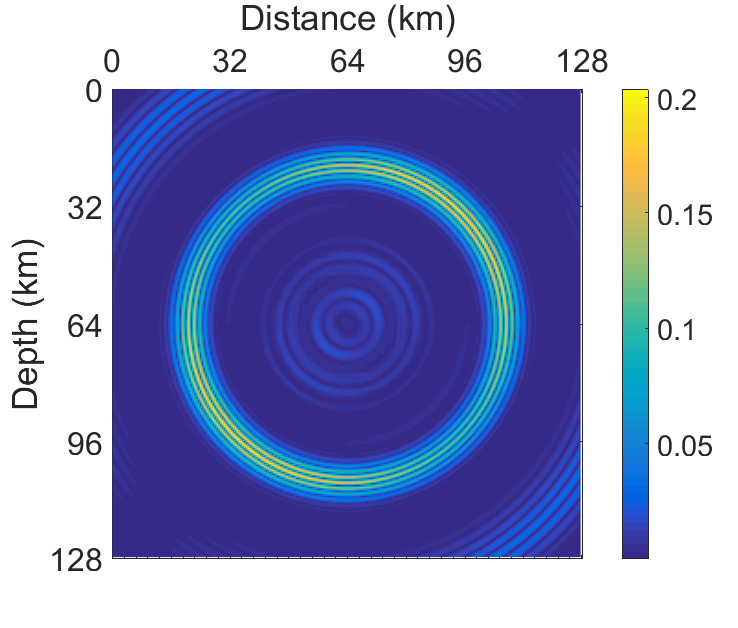}
\end{minipage}}
\subfigure[Solution]
{\begin{minipage}[t]{0.33\linewidth}
\centering
\includegraphics*[scale = .25]{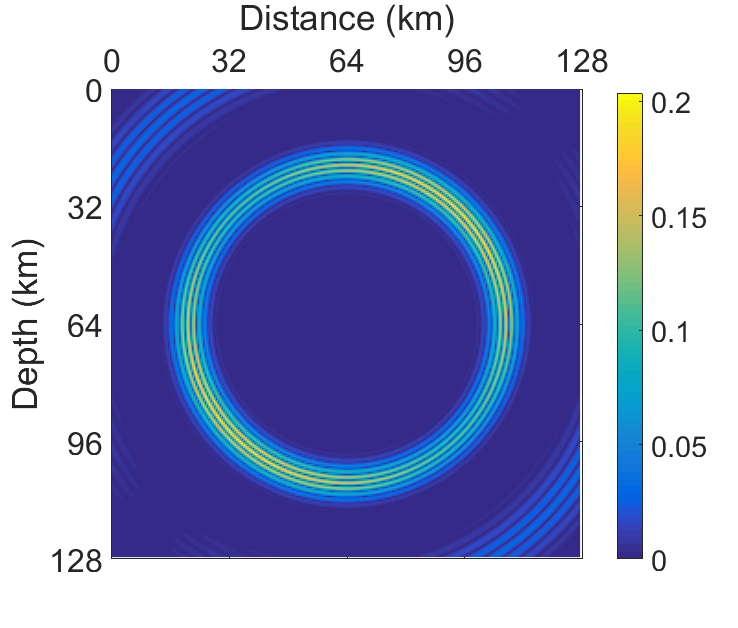}
\end{minipage}}\\
\subfigure[FGA vs Solution]
{\begin{minipage}[t]{0.40\linewidth}
\centering
\includegraphics*[scale = .25]{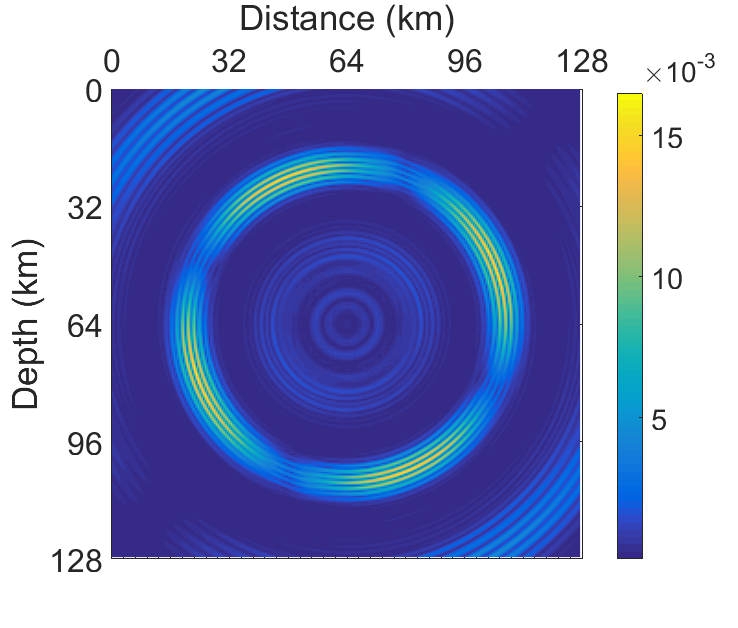}
\end{minipage}}
\subfigure[SEM vs Solution]
{\begin{minipage}[t]{0.40\linewidth}
\centering
\includegraphics*[scale = .25]{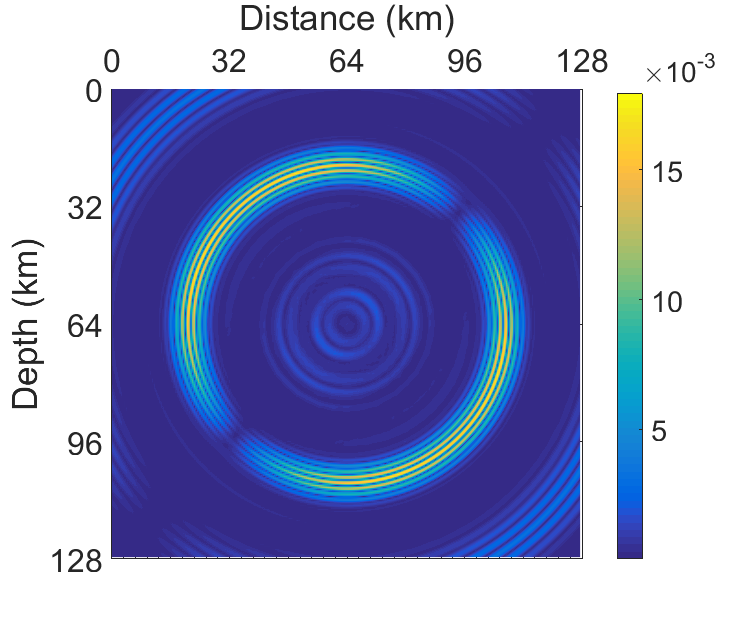}
\end{minipage}}
\caption{Comparison of accuracy for FGA and SPECFEM3D to the analytical solution \eqref{eq:ex_anal} at $t = 6.93$ s with the source frequency $f = 1.4702$ Hz. The top subfigures are the modulus of wavefields given by FGA, SPECFEM and analytical solution. The bottom subfigures are the modulus differences between the FGA and analytical solutions (bottom left) and between the SPECFEM and analytical solutions (bottom right). This shows that FGA and SPECFEM3 produce comparable accuracy in this case.} \label{fig:ex1_FGAvsSEM}
\end{figure*}

\begin{figure}
\centering
\includegraphics*[scale = .35]{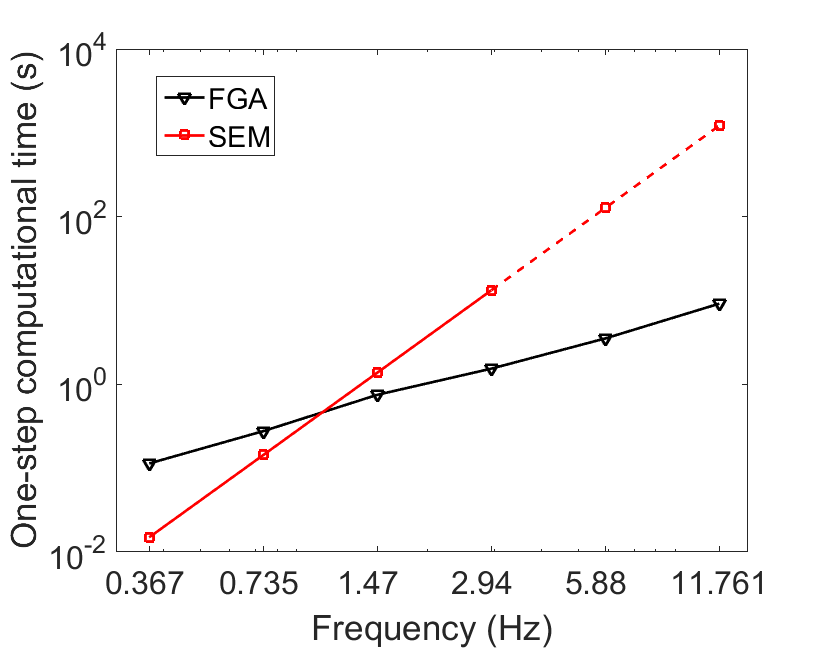}
\caption{Dependence of one-step computational time on frequency for both FGA and SPECFEM3D in homogeneous media. The horizontal axis is the frequency $f$ (Hz) and the vertical axis is the one-step computational time (s) of the solvers. The triangle line stands for the FGA simulations and the square line stands for the SPECFEM3D simulations. Due to the limitation of memory, SPEFEM3D can not run for $f\geq 5.88084$ Hz, and the dashed square line is obtained by extrapolation.}\label{fig:ex1_comptime}
\end{figure}

\begin{figure}
\centering
\includegraphics*[scale = .35]{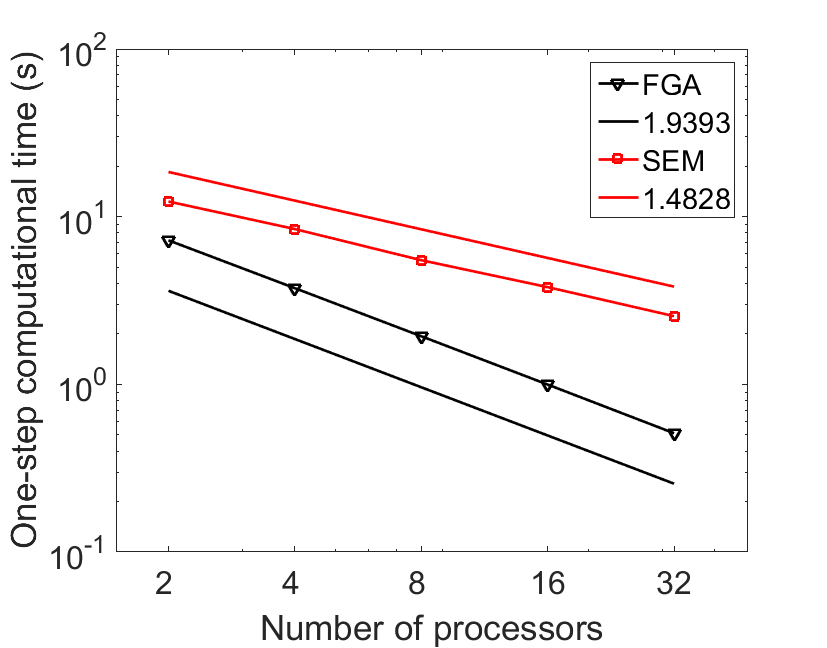}
\caption{Dependence of time on the number of processors for the source frequency $f=1.4702$ Hz in homogeneous media. The ideal speed-up ratio is $2$, while one can see that the speed-up ratio for FGA is approximately $1.9393$, which is slightly smaller than $2$ indicating an almost perfectly parallel efficiency. On the other hand, as a comparison, SPECFEM3D is used with 128 elements in each spatial direction to achieve a comparable accuracy to FGA. The speed-up ratio for the SPECFEM3D solver is around $1.4828$, which is smaller that those of FGA. This is because SPECFEM3D solves eq.~\eqref{eq:ewe} on a parallel computer with $N$ processors by partitioning the whole domain into $N$ slabs with each processor solving the equation in each slab. Therefore, for each time step, each processor needs to communicate with its neighbors to get necessary boundary information, which decreases the speed-up ratio.} \label{fig:ex1_parallel}
\end{figure}

The dynamics of the elastic wave propagation is shown by the FGA method in Fig.~\ref{fig:ex1_dynamics}, which is consistent with the analytical solution \eqref{eq:ex_anal} as an expanding ball. The component-wise wavefields, including both P- and S-wave components, are shown in Fig.~\ref{fig:ex1_components} for $t=6.93$ s. With the comparable accuracy of FGA to SPECFEM3D as illustrated in Fig.~\ref{fig:ex1_FGAvsSEM} for $t = 6.93$ s and the source frequency $f = 1.4702$ Hz.  The relative error for Fig.~\ref{fig:ex1_FGAvsSEM}(d), FGA versus the analytical solution is computed to be 3.84\%; while for Fig.~\ref{fig:ex1_FGAvsSEM}(e), the relative error is computed to be 3.57\%. With a comparable accuracy, the FGA shows a much faster computational speed and better parallelizablity than SPECFEM3D for high-frequency ($\geq 1.4702$ Hz) elastic wave propagation, with details described in Figs.~\ref{fig:ex1_comptime} and~\ref{fig:ex1_parallel}. Particularly, one can see in Fig.~\ref{fig:ex1_comptime} that the computational time of SPECFEM3D has nearly cubic growth in the frequency of elastic waves, while FGA has about $1.5$-times growth in the frequency. Fig.~\ref{fig:ex1_parallel} shows that the speed-up ratio for FGA is almost 2 when one doubles the number of processors, which indicates that the FGA algorithm is embarrassingly parallel. Remark that, in the simulation for the source frequency $f = 1.4702$ Hz where FGA and SPECFEM3D produce comparable accuracy, we use $1504436$ Gaussians for computing P-wave of both ``$\pm$'' propagation directions, and $2120482$ Gaussians for computing S-wave of both ``$\pm$'' propagation directions, which needs to roughly compute a total number of $8\times(1504436+2120482)\approx $ $30$ millions of variables in the simulation. Note that, the equations for both P- and S-waves have the same number of variables, and the prefactor $8$ comes by counting the number of variables needed in eqs.~\eqref{eq:char_plus} and \eqref{eq:amp_P} where $\bd{Q}$ and $\bd{P}$ are 3-D real vectors and the prefactor amplitude $a$ is a complex number; while in SPECFEM3D, we use $128$ elements in each direction with $5^3$ nodes in each element. One needs to roughly to compute a total number of $3\times128^3\times5^3\approx$ $800$ millions of variables in the simulation, where the prefactor $3$ is due to $\bd{u}$ is a 3-D real vector in eq.~\eqref{eq:ewe}. In addition, the stability conditions on time step for solving the ODE systems \eqref{eq:char_plus} and \eqref{eq:amp_P} by RK4 is better than solving the elastic wave equation \eqref{eq:ewe} by SPECFEM, since the CFL condition is restricted by small wavelength in solving eq.~\eqref{eq:ewe}.

\section{Interface conditions and simulation of a 1-D layered Earth model}
In this section, we derive the interface conditions of FGA for 3-D elastic wave propagation in heterogeneous media with strong discontinuities, e.g., the Moho surface and Core-mantle boundary. We verify these interface conditions by simulating a 1-D layered Earth model, which actually requires to revise the FGA solution in spherical coordinates (non-Cartesian coordinates). This can be easily done by an observation that the only step of the FGA algorithm depending on the $\bd{x}$-coordinate is the reconstruction part given by eq.~\eqref{eq:FGA} and all the other steps including the initial wavefield decomposition by eqs.~\eqref{eq:psi} and \eqref{eq:alpha^k} and propagation of ray equations eqs.~\eqref{eq:char_plus}, \eqref{eq:char_minus}, \eqref{eq:amp_P}, \eqref{eq:amp_SV} and \eqref{eq:amp_SH} are $\bd{x}$-coordinate-free. Therefore, to generalize the FGA method in spherical coordinates, one simply needs to change $\bd{x}=(x,y,z)^\TT$ in eq.~\eqref{eq:FGA} to $\bd{x}=(r\cos \theta \cos \phi, r\sin \theta \cos \phi, r\sin \phi)$ with $r$ as the distance to earth center, $\theta$ and $\phi$ as the longitude and latitude degrees.

\subsection{Transmission and Reflection Interface Conditions}\label{sec:TR_cond}
For a sake of clarity and simplicity, we only give the interface conditions for a flat interface located at $z=z_0$, and for a reflecting geometry of general shape, one needs to apply the formulation in the local tangent-normal coordinates by treating the tangential direction as the local flat horizontal interface. The wave speeds of the two layers are assumed to be, 
\begin{align}
c_{\tp}(\bd{x}) =
\begin{cases}
c_{\tp}^\vee(\bd{x}) & z > z_0 \\
c_{\tp}^\wedge(\bd{x}) & z < z_0
\end{cases},\quad
c_{\ts}(\bd{x}) =
\begin{cases}
c_{\ts}^\vee(\bd{x}) & z > z_0 \\
c_{\ts}^\wedge(\bd{x}) & z < z_0
\end{cases}.
\end{align}
 In Fig.~\ref{fig:interface}, we only consider an incident Gaussian wave packet for P-wave hitting the interface at $z=z_0$, and then reflected and transmitted as Gaussian wave packets for P- and SV-waves, respectively. The other cases including an incident Gaussian wave packet for SV- and SH-waves can be handled similarly, although there is no interaction between P- and S-waves for the case of SH-wave.

Associated with each Gaussian wave packet, one needs to provide the reflection and transmission conditions for $a_{\tp,\ts}$, $\bd{Q}_{\tp,\ts}$, and $\bd{P}_{\tp,\ts}$, which follow the Snell's Law and the Zoeppritz equations \citep{Yi:01}. However, for FGA, what is different from standard ray theory on the flat interface is that, one also needs to derive the interface conditions for $Z_{\tp,\ts}$ which will change after the Gaussian wave packet hits the interface and affect the dynamics of $a_{\tp,\ts}$ given by eqs.~\eqref{eq:amp_P}-\eqref{eq:amp_SH}. Eq.~\eqref{eq:op_zZ_app} implies that to derive the interface conditions of $Z_{\tp,\ts}$ will be equivalent to derive the interface conditions for $\partial_{\bd{z}}\bd{Q}_{\tp,\ts}$ and $\partial_{\bd{z}}\bd{P}_{\tp,\ts}$, which requires to use the conservation of level set functions designed in the Eulerian frozen Gaussian approximation formula \citep{LuYa:MMS,WeYa:12}. The mathematical details of the derivation are lengthy and technical, and thus we leave them to Appendix~\ref{app:interface} for the interested readers, and only present the results here:
\begin{equation}\label{eq:Inter_DzQDzP}
\begin{aligned}
    \p_\bz\bQ^{\RE,\TR}&=\p_\bz\bQ^{\IN}\,F,\\
    \p_\bz\bP^{\RE,\TR}&=\p_\bz\bP^{\IN}\,W-\f{\abs{\bP^{\RE,\TR}}}{c(\bQ^{\RE,\TR})p_z^{\RE,\TR}}\left(\p_\bz\bQ^{\RE,\TR}
    \cdot{\nb}c(\bQ^{\RE,\TR})-\p_\bz\bQ^{\IN}\cdot{\nb}c(\bQ^{\IN})\right)\bd{e}_3,
\end{aligned}
\end{equation}
where $(\bQ^{\IN,\RE,\TR},\bP^{\IN,\RE,\TR})$ corresponds to the center and propagation vector of incident, reflected and transmitted Gaussian wave packet for either P- or S-waves, respectively, and $\bP^{\IN,\RE,\TR}=(p_x,p_y,p_z^{\IN,\RE,\TR})$. Here $\bd{e}_3=(0,0,1)$ is a row vector,  $F$ and $W$ are two $3\times3$ matrices, $F^T=W^{-1}$, and
   \begin{align*}
    F=\begin{bmatrix}
    1&&\\&1&\\
    \left(\kappa-1\right)\f{p_x}{p_z^{\IN}}&\left(\kappa-1\right)\f{p_y}{p_z^{\IN}}&\kappa\f{p_z^{\RE,\TR}}{p_z^{\IN}}
      \end{bmatrix},\quad\text{with}\quad \kappa=\left(\frac{c(\bQ^{\RE,\TR})}{c(\bQ^{\IN})}\right)^2.
   \end{align*}

\begin{figure}
\begin{center}
\input{interface_PS.pspdftex}
\end{center}
\caption{Cartoon illustration of an incident Gaussian wave packet for P-wave hitting the interface at $z=z_0$, and then reflected and transmitted as Gaussian wave packets for P- and SV-waves. Here the $G^{\text{in,re,tr}}_{\text{p,s}}$ stands for the Gaussian wave packet for the incident, reflected and transmitted P- and SV-waves, respectively. We denote $\theta_i, \theta_r, \theta_t$ to be the incident, reflection and transmission angles of P-waves, and $\phi_r, \phi_t$ to be the reflection and transmission angles of SV-waves, respectively.}\label{fig:interface}
\end{figure}
Let us explain the formulation of reflection and transmission interface conditions with more detials for the incident P-wave as illustrated in Fig.~\ref{fig:interface}. In this case, if one denotes $(\bQ^{\IN,\RE,\TR}_{\tp,\ts},\bP^{\IN,\RE,\TR}_{\tp,\ts}, a^{\IN,\RE,\TR}_{\tp,\ts})$ to be the center, propagation vector and prefactor amplitude of the corresponding incident, reflected and transmitted Gaussian wave packets for P- and S-waves respectively, and $\bd{P}^{\IN}_{\tp}=(p_x,p_y,p_z)$, then for $\bd{P}^{\RE,\TR}_{\tp}=(p_x,p_y,p^{\RE,\TR}_{z,\tp})$ and $\bd{P}^{\RE,\TR}_{\ts}=(p_x,p_y,p^{\RE,\TR}_{z,\ts})$,
\begin{align} \label{eq:p:int}
p^{\RE}_{z,\tp}&=-p_z, \nonumber\\
p^{\TR}_{z,\tp}&=\sqrt{(p_zn_1)^2+(n_1^2-1)(p_x^2+p_y^2)}, \nonumber\\
p^{\RE}_{z,\ts}&=-\sqrt{(p_zn_2)^2+(n_2^2-1)(p_x^2+p_y^2)}, \\
p^{\TR}_{z,\ts}&=\sqrt{(p_zn_3)^2+(n_3^2-1)(p_x^2+p_y^2)}, \nonumber
\end{align}
where $n_1=c_{\tp}^\wedge/c_{\tp}^\vee$, $n_2=c_{\tp}^\wedge/c_{\ts}^\wedge$ and $n_3=c_{\tp}^\wedge/c_{\ts}^\vee$ with $\displaystyle c_{\tp,\ts}^\vee=c_{\tp,\ts}^\vee(\bd{x})|_{z=z_0}$ and $\displaystyle c_{\tp,\ts}^\wedge=c_{\tp,\ts}^\wedge(\bd{x})|_{z=z_0}$.

Moreover, if one denotes $\theta_i, \theta_r, \theta_t$ to be the P-wave incident, reflection and transmission angles, and $\phi_r, \phi_t$ to be the SV-wave reflection and transmission angles, respectively, then the Zoeppritz equations \citep{Yi:01} read as
\begin{equation}\label{eq:Yilmaz}
M\left(
\begin{array}{c}
a_\tp^{\RE} \\
a_\ts^{\RE} \\
a_\tp^{\TR} \\
a_\ts^{\TR} \\
\end{array}
\right)
=
\left(
\begin{array}{c}
\cos(\theta_r) \\
\sin(\theta_r) \\
\cos(2\phi_r) \\
\cos(2\theta_r) \\
\end{array}
\right)a_\tp^{\IN},
\end{equation}
with the matrix M as
\begin{equation}
 M =
\left(\begin{array}{c c c c}
\cos(\theta_r) & \frac{c_{\tp}^\wedge}{c_{\ts}^\wedge}\sin(\phi_r)      & \frac{c_{\tp}^\wedge}{c_{\tp}^\vee}\cos(\theta_t) & -\frac{c_{\tp}^\wedge}{c_{\ts}^\vee}\sin(\phi_t) \\
-\sin(\theta_r) & \frac{c_{\tp}^\wedge}{c_{\ts}^\wedge}\cos(\phi_r)     & \frac{c_{\tp}^\wedge}{c_{\tp}^\vee}\sin(\theta_t) & \frac{c_{\tp}^\wedge}{c_{\ts}^\vee}\cos(\phi_t) \\
-\cos(2\phi_r) & -\sin(2\phi_r)                      & \frac{\rho_2}{\rho_1}\cos(2\phi_t) & -\frac{\rho_2}{\rho_1}\sin(2\phi_t) \\
\sin(2\theta_r) & -(\frac{c_{\tp}^\wedge}{c_{\ts}^\wedge})^2\cos(2\phi_r)& \frac{\rho_2(c_{\tp}^\wedge c_{\ts}^\vee)^2}{\rho_1(c_{\tp}^\vee c_{\ts}^\wedge)^2}\sin(2\theta_t) & \frac{\rho_2(c_{\tp}^\wedge)^2}{\rho_1(c_{\ts}^\wedge)^2}\cos(2\phi_t)
\end{array}
\right),
\end{equation}
where $\rho_{1,2}$ are the densities for the layers 1 and 2, respectively.

More explicitly in eq.~\eqref{eq:Inter_DzQDzP} corresponding to the case in Fig.~\ref{fig:interface}, $c(\bd{Q}_{\tp}^\IN)=c_{\tp}^\wedge(\bd{Q}_{\tp}^\IN)$, $c(\bd{Q}_{\tp}^\RE)=c_{\tp}^\wedge(\bd{Q}_{\tp}^\RE)$, $c(\bd{Q}_{\tp}^\TR)=c_{\tp}^\vee(\bd{Q}_{\tp}^\TR)$, $c(\bd{Q}_{\ts}^\RE)=c_{\ts}^\wedge(\bd{Q}_{\ts}^\RE)$, and $c(\bd{Q}_{\ts}^\TR)=c_{\ts}^\vee(\bd{Q}_{\ts}^\TR)$.

\subsection{Waveguide example in a 1-D layered Earth model}
\begin{figure}
\subfigure[]{
\begin{minipage}{0.5\textwidth}
 \begin{center}
  {\includegraphics[scale=0.5]{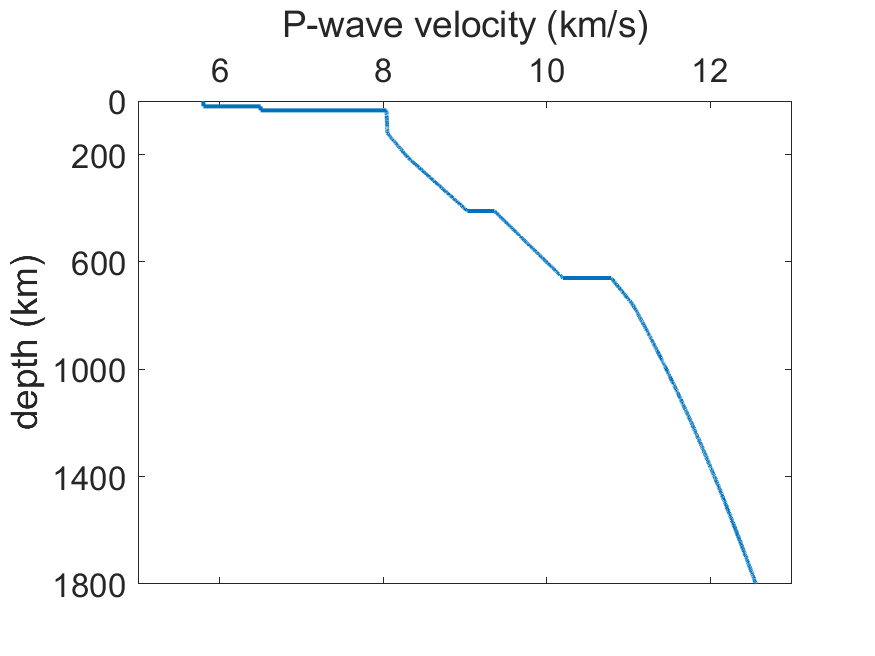}}
 \end{center}
\end{minipage}}
\subfigure[]{
\begin{minipage}{0.5\textwidth}
 \begin{center}
  {\includegraphics[scale=0.5]{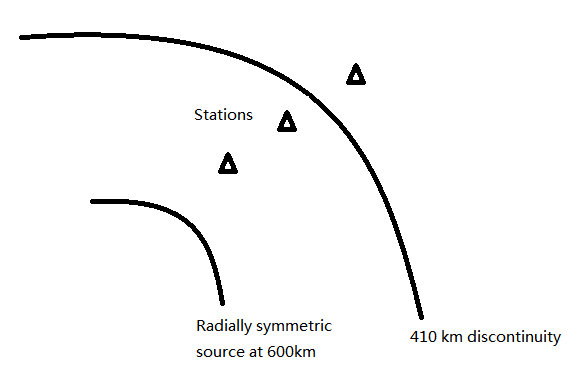}}
 \end{center}
\end{minipage}}
 \caption{(a): The layered P-wave velocity follows the data in the IASP91 model \citep{Ke:91,DMC:10}. (b): Cartoon plot of the $410$-km discontinuity, and distributions of source and stations.}
 \label{fig:iasp91_setup}
\end{figure}

We verify the interface conditions \eqref{eq:Inter_DzQDzP} and \eqref{eq:Yilmaz} by simulating a waveguide example in a 1-D layered Earth model, with the layered P-wave velocity given in Fig.~\ref{fig:iasp91_setup}(a) following the data in the IASP91 model \citep{Ke:91,DMC:10}. We are particularly interested in choosing the $410$-km discontinuity for the numerical proof of the conditions \eqref{eq:Inter_DzQDzP} and \eqref{eq:Yilmaz}, which presents to a $5-6\%$ increase on P-wave velocity and calibrates the mantle transition zone. We consider a radially symmetric surface source as shown in Fig.~\ref{fig:iasp91_setup}(b), so that the elastic wave equation \eqref{eq:ewe} has a solution of P-waves in the form of $\bu(t,\bx)=\grdx\psi(t,\abs{\bx})$ where $\psi(t,r)$ is the radially symmetric solution to the scalar wave equation, i.e.,
\begin{equation}\label{eq:we_radially_symmetric}
\partial_t^2\psi-c_{\tp}^2(\bd{x})\Bigl(\partial_r^2 \psi + \frac{2}{r} \partial_r \psi\Bigr)=0.
\end{equation}

\begin{figure}
\subfigure[P-wave at depth $480$ km]
{\begin{minipage}[t]{0.33\linewidth}
\centering
\includegraphics*[scale = .36]{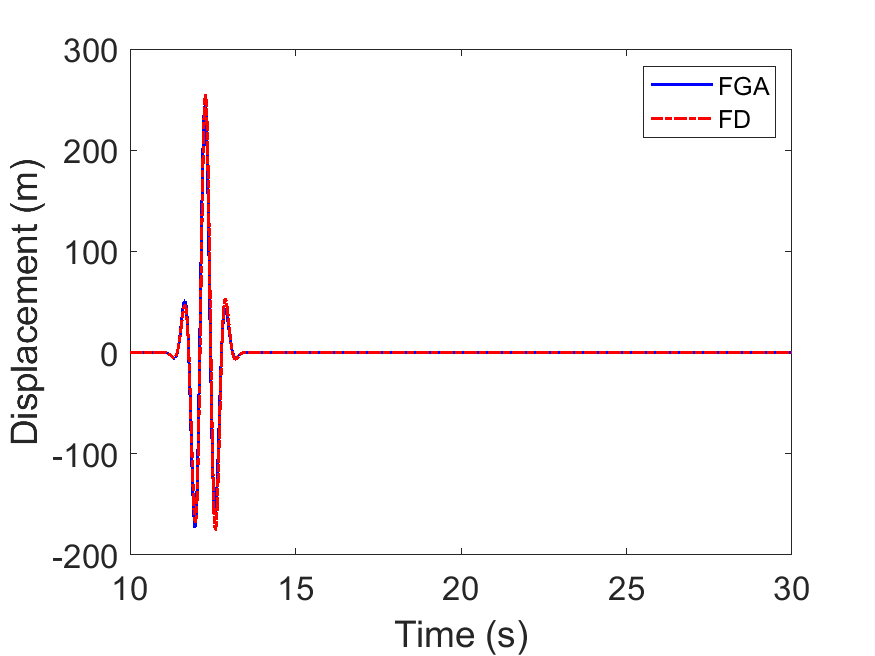}
\end{minipage}}
\subfigure[P-wave at depth $420$ km]
{\begin{minipage}[t]{0.33\linewidth}
\centering
\includegraphics*[scale = .36]{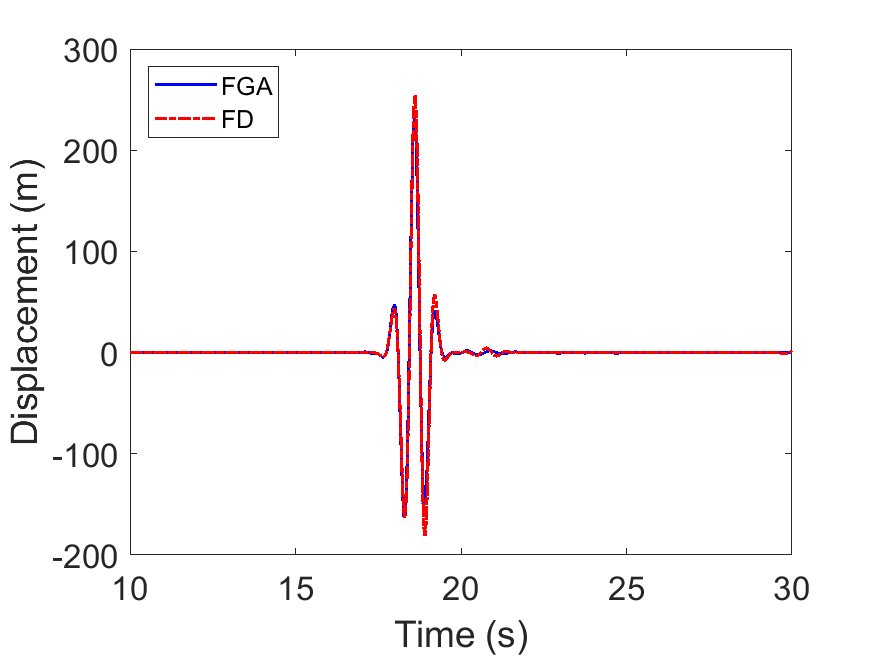}
\end{minipage}}
\subfigure[P-wave at depth $360$ km]
{\begin{minipage}[t]{0.33\linewidth}
\centering
\includegraphics*[scale = .36]{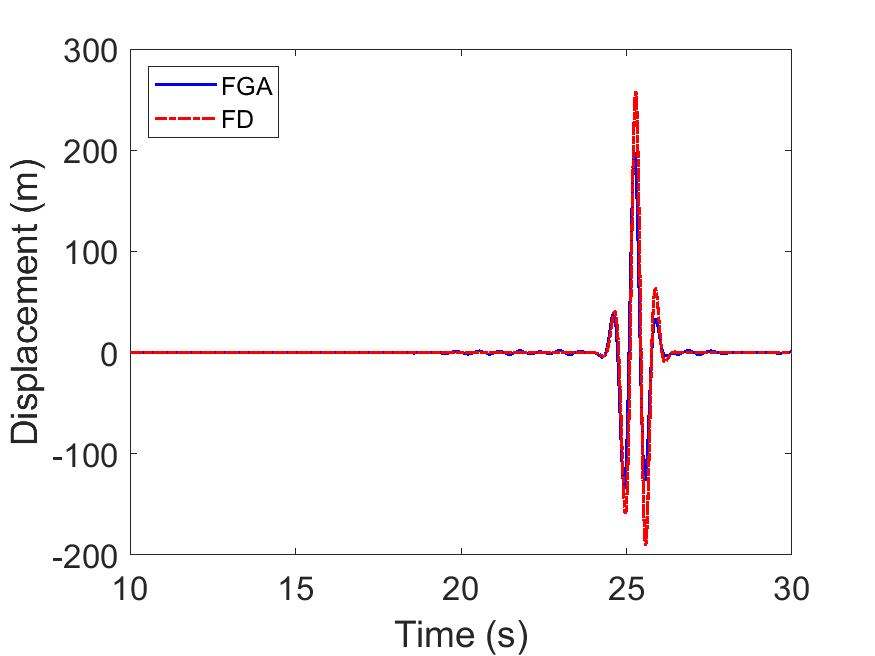}
\end{minipage}}
\caption{The seismic signals of P-waves received at stations of depth $480$ km, $420$ km and $360$ km, respectively, simulated by the finite difference (FD) and FGA methods, with the source locating at the depth of $600$ km. FGA shows a good agreement with the reference signals computed by the FD method for the P-waves.}\label{fig:iasp91}
\end{figure}

We choose the initial condition for the elastic wave equation \eqref{eq:ewe} as
\begin{equation*}
 \bu(t=0,\bx)=\grdx\phi_0(r-r_0),\;\text{ and }\;
 \p_t\bu(t=0,\bx)=\bd0,
\end{equation*}
where $r=\abs{\bx}$, $r_0=600$ km, and
\begin{equation*}
 \phi_0(r) = \exp\left(-\f{r^2}{2\sg^2}\right)\cos\left(\f{2\pi r}{\ell}\right),
\end{equation*}
with $\sg=3.3146$ km and $\ell=7.3631$ km. The dominant frequency is around $1.36$ Hz.
We solve eq.~\eqref{eq:we_radially_symmetric} using 1-D finite difference method with fine enough grid points as the reference solution, to which we compare the full elastic wave solution of \eqref{eq:ewe} computed by the FGA algorithm. Fig.~\ref{fig:iasp91} shows the seismic signals of P-waves received at stations of depth $480$ km, $420$ km and $360$ km, respectively, where one can see a good agreement of FGA simulation with the reference solution for the P-waves.

\section{{ Wave-equation-based Traveltime Tomography} and FWI}
\begin{figure}
\subfigure[Three-layered crosswell model]{
\begin{minipage}{0.5\textwidth}
 \begin{center}
  {\includegraphics[scale=0.9]{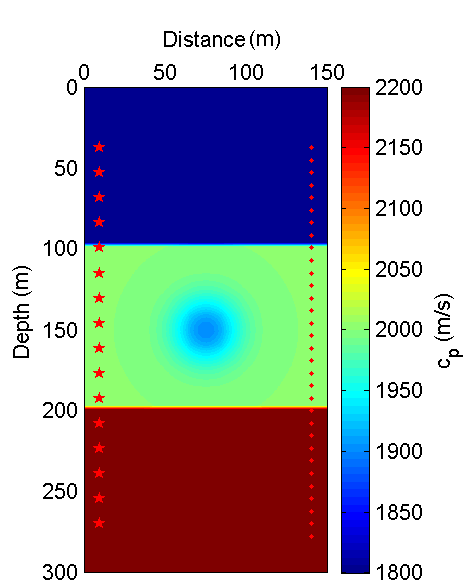}}
 \end{center}
\end{minipage}}
\subfigure[Relative velocity $(c_\tp-c_0)/{c_0}$]{
\begin{minipage}{0.5\textwidth}
 \begin{center}
  {\includegraphics[scale=0.9]{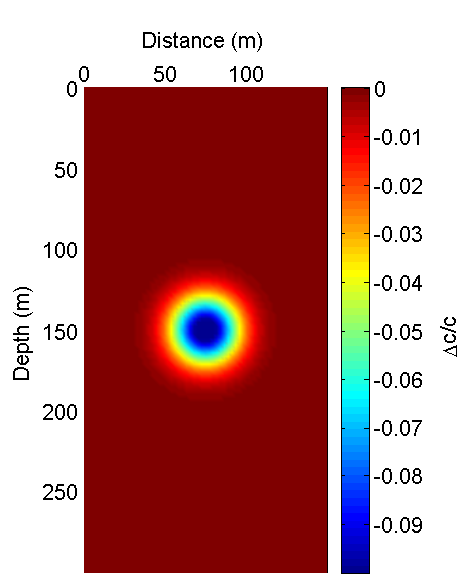}}
 \end{center}
\end{minipage}}
 \caption{(a): The three-layered crosswell P-wave velocity model with homogeneity in horizontal $y$-direction. The background velocities from top to bottom are $C_1=1800$ m/s, $C_2=2000$ m/s, $C_3=2200$ m/s, and the interfaces locate at $z_0=0$ m, $z_1=100$ m, $z_2=200$ m. The low-velocity region has a Gaussian shape centered at $x_c=75$ m, $z_c=150$ m, with standard deviation equal to $1/30$ m. $16$ stars indicate the locations of seismic sources, and $32$ dots indicate the locations of seismic receivers. (b): we use the relative velocity $\Delta c/c=(c_\tp-c_0)/c_0$ to indicate the largest magnitude of the low-velocity perturbation from the background velocity and the area of low velocity region.}
 \label{fig:cross-well_setup}
\end{figure}

In this section, we apply the developed FGA algorithm for 3-D { wave-equation-based traveltime tomography \citep[e.g.][]{Tromp2005,Liu2012} and FWI \citep[e.g.][]{Pratt1999,Virieux2009}}, respectively. In particular, we consider the following crosswell seismic tomography, where the three-layered velocity model is set up as follows, with a low-velocity region located at the second layer and homogeneity in the $y$-direction,
\begin{equation}\label{eq:true_c}
 c_{\tp}(x,y,z)=\sqrt{3}c_{\ts}(x,y,z)=
 \begin{dcases}
  C_1, & \text{ if } z_0 < z < z_1, \\
  C_2\left(1-\alpha\e^{-\beta\left((x-x_c)^2+(z-z_c)^2\right)}\right), & \text{ if } z_1 < z < z_2,\\
  C_3, & \text{ if } z > z_2,
 \end{dcases}
\end{equation}
where the layered velocities are $C_1=1800$ m/s, $C_2=2000$ m/s, $C_3=2200$ m/s, and the interfaces locate at $z_0=0$ m, $z_1=100$ m, $z_2=200$ m. The center of the low-velocity region is set at $x_c=75$ m, $z_c=150$ m. We choose $\alpha=10\%$ and $\beta=1/450\text{ m}^{-2}$ to indicate the largest magnitude of the low-velocity perturbation from the background velocity and the area of low velocity region;
see Fig.~\ref{fig:cross-well_setup} for an illustration of the P-wave velocity. In Fig.~\ref{fig:cross-well_setup}, we also show the positions of 16 seismic resources (as stars) with an equal spacing of $16$ m in one well and 32 seismic receivers (as dots) with an equal spacing of $8$ m in the other well.


\begin{figure}
\centering
\includegraphics*[scale = .5]{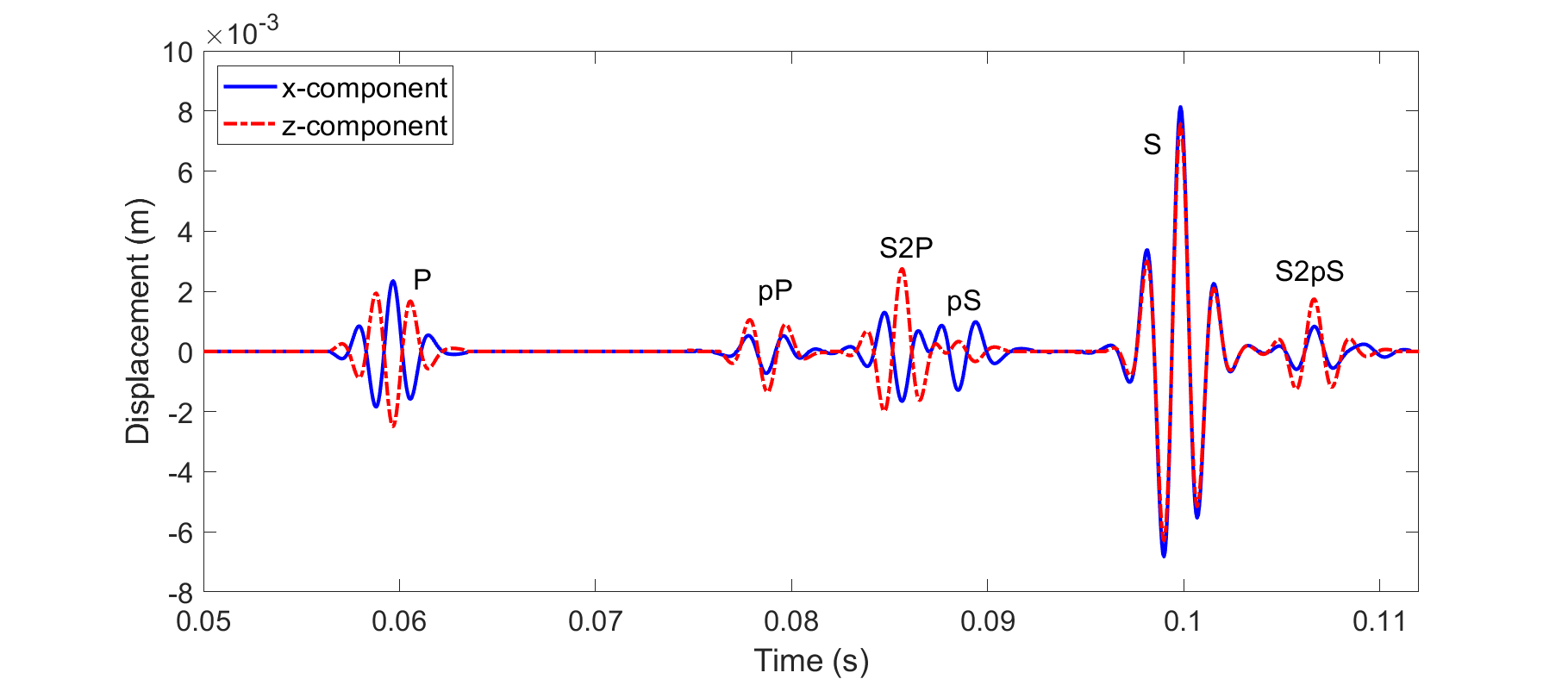}
\caption{The signals of direct P, pP, S2P, pS, direct S, and S2pS received at the top station with the source as the top ninth source in Fig.~\ref{fig:cross-well_setup}. Here $2$ means the second interface locates at $z_2=200$ m, e.g. S2P means S-wave gets reflected at the interface at $z_2=200$ m and arrives at the station as P-wave. }\label{fig:signals}
\end{figure}

\begin{figure}
\includegraphics*[scale = 0.9]{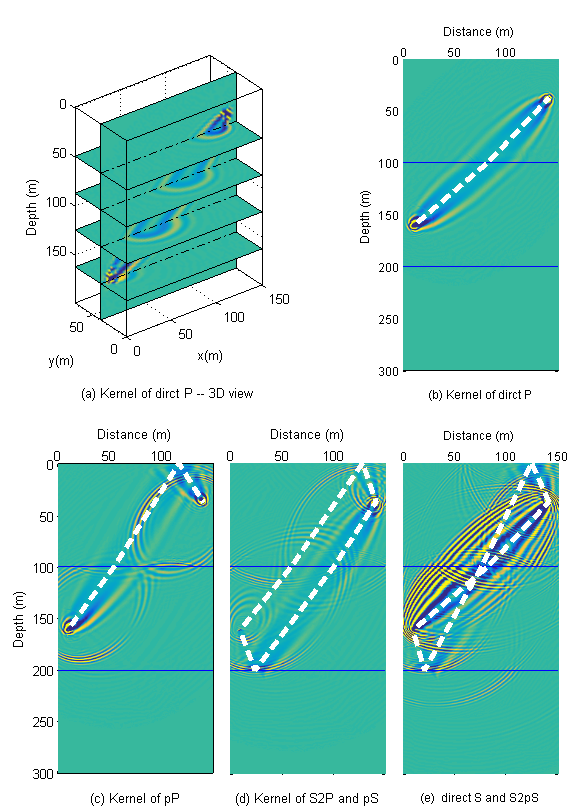}
 \caption{Seismic tomography kernels computed by FGA for the initial velocity model \eqref{eq:ini_c}, with the thick dashed lines as the actual ray paths of direct P, pP, S2P and pS, direct S, and S2pS signals in Fig.~\ref{fig:signals}, respectively.
 { (a): Kernel computed from the direct P signal -- 3D slices view};
 (b): Kernel computed from the direct P signal;
 (c): Kernel computed from the pP signal;
 (d): Kernel computed from the S2P and pS signals;
 (e): Kernel computed from the direct S and S2pS signals.
 }
 \label{fig:kernel}
\end{figure}

\begin{figure}
\centering
\includegraphics*[scale = 0.9]{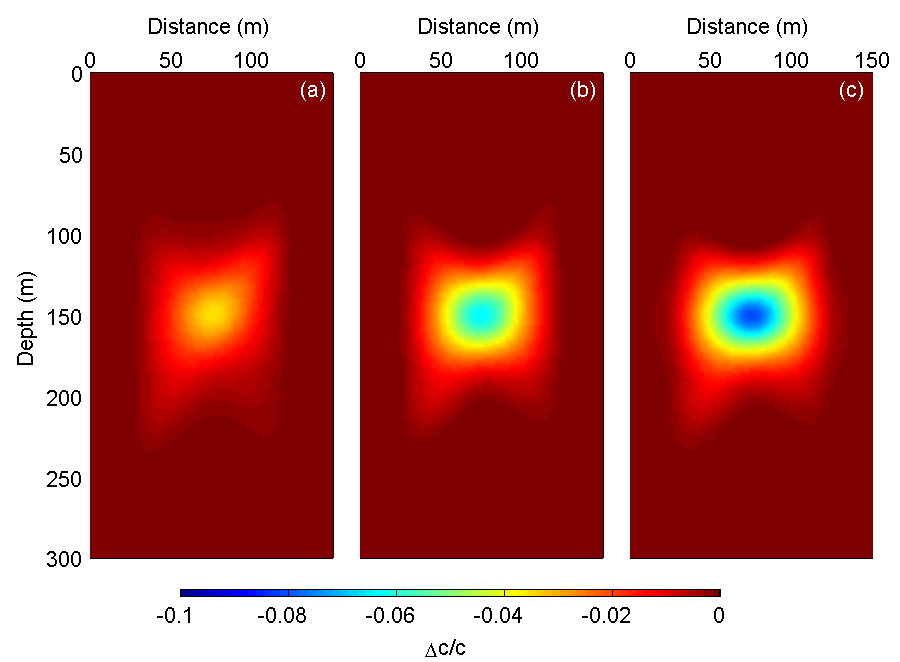}
\caption{For $\alpha=10\%$ and $\beta=1/450\text{ m}^{-2}$ in eq.~\eqref{eq:true_c}, the subfigures (a) and (b) are the first two iterations using travel-time tomography, and the subfigure (c) is the third iteration using FWI which removes the artifacts in travel-time tomography and is close to the true velocity profile given in Fig.~\ref{fig:cross-well_setup}(b). }\label{fig:cross-well-inv}
\end{figure}

We start with the following initial velocity model
\begin{equation}\label{eq:ini_c}
 c_0(x,y,z)=
 \begin{dcases}
  C_1, & \text{ if } z_0 < z < z_1, \\
  C_2, & \text{ if } z_1 < z < z_2,\\
  C_3, & \text{ if } z > z_2,
 \end{dcases}
\end{equation}
then use FGA to compute the forward and adjoint wavefields and construct 3-D kernels of different phases by the methods of { wave-equation-based traveltime tomography} \citep{Tromp2005,Liu2012} and FWI \citep{Pratt1999,Virieux2009}. Since FWI requires a more sophisticated initial velocity model for the convergence than {wave-equation-based} traveltime tomography, we use a hierarchical strategy as in our previous work \citep{chai2018tomo}, which first uses { wave-equation-based traveltime tomography} to create a macro-scale model and then adopts FWI to generate a high-resolution micro-scale model. For the signals received at the top station (Fig.~\ref{fig:signals}), we show the corresponding kernels of different phases in Fig.~\ref{fig:kernel}. The inversion results are shown in Fig.~\ref{fig:cross-well-inv}. The damping and smoothing parameters are chosen empirically by forcing the variation in each iteration less than 3\%. Note that artifacts are visible in the final images, which are unavoidable and mainly caused by the uneven data coverages.

\section{Discussion and conclusions}
This is the last part of our theoretical study on the application of the frozen Gaussian approximation (FGA) method to seismic tomography using high-frequency seismic data. Together with \citet{chai2017frozen,chai2018tomo}, we systematically derive the FGA formulation for acoustic and elastic wave equations, derive the transmission and reflection conditions of FGA for sharp interfaces (e.g., Moho surface and Core-mantle boundary), reformulate the equations of the forward and adjoint wavefields for a convenient application of the FGA algorithm, propose a fast Gaussian summation algorithm for the reconstruction of FGA, and {conduct 3-D high-frequency seismic inversion} by the methods of { wave-equation-based traveltime tomography and FWI}. As a proof of methodology, we compare FGA to SPECFEM3D for acoustic and elastic wave propagation in homogeneous media where an analytical benchmark solution is available. We also test the performance of FGA in a 1-D layered Earth model for checking the accuracy of the interface conditions, and apply FGA to the cross-well example for the 3-D seismic tomography using high-frequency data. We remark that, in the original mathematical work on FGA for strictly linear hyperbolic systems \citep{LuYa:CPAM}, the FGA formulation is rather complicated, and done for a purpose of rigorously proving its accuracy in the high-frequency regime. In this paper, we weakly expand the solution ansatz of FGA with a consideration of decomposition into P- and S-waves, and reach simple prefactor equations \eqref{eq:amp_P}, \eqref{eq:amp_SV} and \eqref{eq:amp_SH} after lengthy and tricky simplifications. This also generalizes the results in \citet{LuYa:CPAM} to some extent in the sense that the elastic wave equation is actually not strictly hyperbolic, which has an eigenfunction space of multiplicity-2 (corresponding to S-waves). By these efforts, we hope to bridge a gap between the semiclassical approximation theory in mathematics and the application of 3-D seismic tomography in geophysics. { We also remark that, it will be interesting to study the performance of the FGA method in the optimal transport theory-based
seismic tomography, with the normalization strategies proposed in, e.g., \citet{EnFr:14,MeBrMeOuVi:16,qiu2017full,YaEnSuHa:18,ChChWuYa:arXiv}.} Our future plans include: 1. to apply 3-D seismic tomography with the FGA method into real data around their dominant frequencies; 2. to use the fast computation performance of FGA to train neural networks for seismic inversion; { 3. to develop the FGA algorithm for the optimal transport theory-based seismic tomography using high-frequency data}.

\begin{acknowledgments}
We are grateful to the support from the Center for Scientific Computing from the CNSI, MRL: an NSF MRSEC (DMR-1121053) and NSF CNS-0960316. JH, LC and XY were partially supported by the NSF grants DMS-1818592, DMS-1418936 and DMS-1107291. JH was also partially supported by the graduate summer fellowship of Department of Mathematics at University of California, Santa Barbara. PT was partially supported by the MOE AcRF Tier-1 grant (M4011899.110). XY also thanks Professor Qinya Liu for useful discussions on the WKM and GRT methods. We also thank the anonymous referee for bringing us the interesting questions on the optimal transport theory-based seismic tomography.
\end{acknowledgments}

\appendix

\section{Derivation of the FGA formulation}\label{app:FGA}
This appendix is devoted to the mathematical details of deriving the FGA formulation. It is lengthy and technically involved, making use of a few mathematical concepts including weak asymptotic expansion, Schwartz class functions, canonical transform and symplecticity. We shall do our best to make the details as straightforward as possible so that they are accessible for interested readers with different background.

For a sake of clarity and simplicity, we only consider the case of $\lambda$ and $\mu$ being constants in eq.~\eqref{eq:ewe} with only ``$+$'' wave propagation direction, and the derivation of FGA for general $\lambda(\bd{x})$ and $\mu(\bd{x})$ with two-way wave propagation directions will be essentially the same but just with more calculations. We plug eq.~\eqref{eq:psi} into eq.~\eqref{eq:FGA}, combine the phase functions, and define the total phase $\Phi_{\tp,\ts}$ as
\begin{equation}\label{eq:phase_func}
\Phi_{\tp,\ts}(t,\bd{x},\bd{y},\bd{q},\bd{p}) = \bd{P}_{\tp,\ts}\cdot(\bd{x} - \bd{Q}_{\tp,\ts}) - \frac{1}{2} \abs{\bd{x} - \bd{Q}_{\tp,\ts}}^2 - \bd{p}\cdot(\bd{y} - \bd{q}) - \frac{1}{2} \abs{\bd{y} - \bd{q}}^2.
\end{equation}
We call two functions $f(\bd{y},\bd{q},\bd{p})$ and $g(\bd{y},\bd{q},\bd{p})$ are weakly equal to each other, denoted by $f\sim g$, if
  \begin{equation}\label{eq:sim}
     \iiint f(\bd{y},\bd{q},\bd{p})\exp(\I k\Phi_{\tp,\ts})\ d\bd{y}\ d\bd{q}\ d \bd{p} = \iiint g(\bd{y},\bd{q},\bd{p})\exp(\I k\Phi_{\tp,\ts})\ d\bd{y}\ d\bd{q}\ d \bd{p}.
\end{equation}
Note that both $f$ and $g$ can be scalar-, vector-, matrix- or tensor-valued functions.

A map: $(\bd{q},\bd{p}) \rightarrow \bigl(\bd{Q}_{\tp,\ts}(\bd{q},\bd{p}),\bd{P}_{\tp,\ts}(\bd{q},\bd{p})\bigr)$ is called canonical transformation if the associated Jacobian matrix
  \begin{equation*}
    J_{\tp,\ts}(\bd{q},\bd{p}) =
    \begin{pmatrix}
      (\partial_q \bd{Q}_{\tp,\ts})^{\TT}(\bd{q},\bd{p}) & (\partial_p
      \bd{Q}_{\tp,\ts})^{\TT}(\bd{q},\bd{p})
      \\
      (\partial_q \bd{P}_{\tp,\ts})^{\TT}(\bd{q},\bd{p}) & (\partial_p
      \bd{P}_{\tp,\ts})^{\TT}(\bd{q},\bd{p})
    \end{pmatrix},
  \end{equation*}
is symplectic, i.e., for any $(\bd{q},\bd{p})$,
   \begin{equation}\label{eq:symplectic}
    J_{\tp,\ts}^{\TT}
    \begin{pmatrix}
      0 & \Id_3 \\
      -\Id_3 & 0
    \end{pmatrix}
    J_{\tp,\ts} =
    \begin{pmatrix}
      0 & \Id_3 \\
      -\Id_3 & 0
    \end{pmatrix},
  \end{equation}
where $\Id_3$ is a $3\times 3$ identity matrix. It is easy to verify that the Hamiltonian flow given by either eq.~\eqref{eq:char_plus} or \eqref{eq:char_minus} is a canonical transform by observing that
   \begin{equation*}
    \frac{\ud}{\ud t} J_{\tp,\ts}(\bd{q},\bd{p})
    =
    \begin{pmatrix}
      \partial_{\bd{P}} \partial_{\bd{Q}} H_{\tp,\ts} & \partial_{\bd{P}} \partial_{\bd{P}} H_{\tp,\ts} \\
      - \partial_{\bd{Q}}p \partial_{\bd{Q}} H_{\tp,\ts} & - \partial_{\bd{Q}} \partial_{\bd{P}} H_{\tp,\ts}
    \end{pmatrix}
    J_{\tp,\ts}(\bd{q},\bd{p}),
  \end{equation*}
with the Hamiltonian function $H_{\tp,\ts}(\bd{Q},\bd{P})= c_{\tp,\ts}(\bd{Q}_{\tp,\ts})\abs{\bd{P}_{\tp,\ts}}$, thus $J_{\tp,\ts}(\bd{q},\bd{p})$ satisfies eq.~\eqref{eq:symplectic}. This property guarantees that the matrix $Z_{\tp,\ts}$ defined in eq.~\eqref{eq:op_zZ_app} is always invertible for all $t>0$ by the following argument, with the subscripts $\tp,\ts$ omitted for convenience,
  \begin{equation*}
    Z{(\bd{q},\bd{p})} = \partial_{z}\bigl(Q{(\bd{q},\bd{p})}
    + \I P{(\bd{q},\bd{p})} \bigr)
    =
    \begin{pmatrix}
      \I \Id_3 & \Id_3
    \end{pmatrix}
    J^{\TT}{(\bd{q},\bd{p})}
    \begin{pmatrix}
      - \I \Id_3 \\
      \Id_3
    \end{pmatrix},
  \end{equation*}
which implies
  \begin{equation*}
    \begin{aligned}
      (Z Z^{\ast}){(\bd{q},\bd{p})} & =
      \begin{pmatrix}
        \I \Id_3 & \Id_3
      \end{pmatrix}
      J^{\TT}{(\bd{q},\bd{p})}
      \begin{pmatrix}
        \Id_3  & -\I \Id_3 \\
        \I \Id_3 & \Id_3
      \end{pmatrix}
      J{(\bd{q},\bd{p})}
      \begin{pmatrix}
        -\I \Id_3 \\
        \Id_3
      \end{pmatrix} \\
      & =
      \begin{pmatrix}
        \I \Id_3 & \Id_3
      \end{pmatrix}
      \bigl(J^{\TT} J\bigr){(\bd{q},\bd{p})}
      \begin{pmatrix}
        - \I \Id_3 \\
        \Id_3
      \end{pmatrix} \\
      & \quad +
      \begin{pmatrix}
        \I \Id_3 & \Id_3
      \end{pmatrix}
      J^{\TT}{(\bd{q},\bd{p})}
      \begin{pmatrix}
        0 & -\I \Id_3 \\
        \I \Id_3 & 0
      \end{pmatrix}
      J{(\bd{q},\bd{p})}
      \begin{pmatrix}
        -\I \Id_3 \\
        \Id_3
      \end{pmatrix}  \\
      & =
      \begin{pmatrix}
        \I \Id_3 & \Id_3
      \end{pmatrix}
      \bigl(J^{\TT} J\bigr){(\bd{q},\bd{p})}
      \begin{pmatrix}
        -\I \Id_3 \\
        \Id_3
      \end{pmatrix} + 2\Id_3,
    \end{aligned}
  \end{equation*}
where $Z^{\ast}$ is the conjugate transpose of $Z$, and in the last equality, we have used eq.~\eqref{eq:symplectic}. Therefore, the $Z^{-1}_{\tp,\ts}$ used in eqs.~\eqref{eq:amp_P}, \eqref{eq:amp_SV} and \eqref{eq:amp_SH} is a well behaved quantity.

With the theoretical preparation above, we are ready to derive the formulation of FGA. We only consider the case $\bd{F}=0$ in eq.~\eqref{eq:ewe}, and refer to \citet[Section 3.1]{chai2018tomo} for the treatment of a general source time function. We start with the following form of Gaussian wave packet as the solution ansatz,
\begin{equation}\label{solform:1}
\bu_{\tp,\ts}(t,\bd{x},\bd{y},\bd{q},\bd{p}) = \bd{A}_{\tp,\ts}(t,\bd{q},\bd{p})\exp\bigl(\I k \Phi_{\tp,\ts}(t,\bd{x},\bd{y},\bd{q},\bd{p})\bigr),
\end{equation}
with $\Phi$ given by eq.~\eqref{eq:phase_func}. Remark that, for the readers who are familiar with Gaussian beam method \citep[e.g.][]{Hi:90,Hi:01,SEG-2003-11141117,Gr:05,GrBe:09,PoSePoVe:10}, it is easy to see that, a direct plugin of eq.~\eqref{solform:1} into eq.~\eqref{eq:ewe} will yield more equations than the number of variables due to the missing of a time-dependent Hessian function in the total phase $\Phi$. Therefore, one can only derive the FGA formulation using the weak asymptotic expansion with a sense of equality defined in eq.~\eqref{eq:sim}, where the following integration by parts lemma will help to eliminate the extra constraints by converting the powers of $\bd{x}-\bd{Q}$ to the powers of $k^{-1}$.

\noindent\textbf{Lemma of integration by parts:} For any vector $\bd{a}(\by,\bq,\bp) = (a_j)$ and matrix $M(\by,\bq,\bp) = (M_{ij})$ in Schwartz class, i.e., with decay properties at infinity so that the integrals in eq.~\eqref{eq:sim} are well defined, one has the following integration by parts formula in the componentwise form, with $\partial_{\bd{z}}=(\partial_{z_1},\partial_{z_2},\partial_{z_3})$,
\begin{equation}\label{lemma:1}
\begin{aligned}
a_j(x-Q)_j \sim&-k^{-1} \partial_{z_m}\big(a_j Z_{jm}^{-1} \big), \\
(x-Q)_jM_{jl}(x-Q)_l \sim &k^{-1} \partial_{z_m}Q_j M_{jl} Z_{lm}^{-1} + \cO(k^{-2}),
\end{aligned}
\end{equation}
where the Einstein's index summation convention is used, the matrix $Z$ is given in eq.~\eqref{eq:op_zZ_app}, and we refer to \citet{LuYa:11,LuYa:CPAM} for the detailed proofs.



\noindent Note that eq.~\eqref{eq:ewe} with $\bd{F}=0$ can be rewritten as the following ``curl'' form,
\begin{equation}\label{eq:ewe_curl}
\rho(\bd{x})\partial^2_t\bu = (\lambda + 2\mu)\nabla(\nabla\cdot \bu) - \mu\nabla\times\nabla \times \bu.
\end{equation}
Notice that eq.~\eqref{eq:ewe_curl} is linear, and thus one can derive the prefactor equations for P- and S-waves individually by assuming $\bd{A}_\tp\parallel \bd{P}$ or $\bd{A}_\ts\perp \bd{P}$ in eq.~\eqref{solform:1}, respectively. Without loss of generality, we first consider the prefactor equation for the P-wave, with the governing equations for $\bd{Q}_\tp$ and $\bd{P}_\tp$ given by eq.~\eqref{eq:char_plus}. For an ease of notations, we shall omit the subscript $\tp$ in calculating the prefactor equation for the P-wave.

\noindent Plugging eq.~\eqref{solform:1} into eq.~\eqref{eq:ewe_curl} and expanding the asymptotics in the weak sense of \eqref{eq:sim} yield
\begin{multline}\label{weq:1}
\rho \left(\bd{A}_{tt} +  2\I k\bd{A}_t\Phi_t +  \I k\bd{A} \Phi_{tt}- k^2 \bd{A} \Phi_t^2  \right)
\sim   (\lambda + 2\mu)\left(\I k\nabla(\bd{A}\cdot \nabla \Phi) -  k^2(\bd{A} \cdot\nabla\Phi)\nabla\Phi \right)  \\ -\mu\left(\I k  \Curl(\nabla \Phi \times \bd{A}) - k^2 \nabla \Phi\times(\nabla \Phi\times \bd{A})\right). \end{multline}
The spatial and temporal derivatives of $\Phi$ are given by
\begin{equation}\label{eq:Phi_der}
\begin{aligned}
&\nabla \Phi =  \bP + \I (\bd{x} - \bQ),\ \Delta \Phi  =  3\I,\ \nabla^2\Phi = \I\Id_3, \\
&|\nabla\Phi|^2 =  |\bP|^2 + 2\I \bP\cdot (\bd{x} - \bQ) - |\bd{x}-\bQ|^2, \\
&\Phi_t =\ (\bP_t- \I \bQ_t)\cdot (\bd{x}-\bQ) - \bP\cdot \bQ_t, \\
&\Phi_t^2 =\ \big[(\bP_t- \I \bQ_t)\cdot (\bd{x}-\bQ)\big]^2 + (\bP\cdot \bQ_t)^2 - 2(\bP\cdot \bQ_t)\big[(\bP_t- \I \bQ_t)\cdot (\bd{x}-\bQ)\big],  \\
&\Phi_{tt} =\ (\bP_{tt}- \I \bQ_{tt})\cdot (\bd{x}-\bQ) - (\bP_t- \I \bQ_t)\cdot \bQ_t - \bP_t\cdot \bQ_t -  \bP\cdot \bQ_{tt}.
\end{aligned}
\end{equation}
Notice that the terms containing $k(\bd{x}-\bd{Q})$ will be of $\cO(1)$ by the lemma of integration by parts, and for P-waves, $\bd{P}\times \bd{A} = 0$ and $\Curl\bigl((\bd{x}-\bQ)\times {\bd{A}}\bigr) = -2\bd{A}$. Plugging the derivatives of $\Phi$ in eq.~\eqref{eq:Phi_der} into eq.~\eqref{weq:1} produces, after neglecting the $\cO(1)$ and lower order terms,
\begin{equation}\label{FGA:1}
\begin{aligned}
2 k \rho \bd{A}_t\big(\bP\cdot \bQ_t \big) \sim &\  k\rho \bd{A} \Big( -(\bP_t- \I \bQ_t)\cdot \bQ_t - \bP_t\cdot \bQ_t -  \bP\cdot \bQ_{tt} \Big) \\
+ &\ \I k \rho \bd{A}\Big( (\bd{x}-\bQ)\cdot\Big((\bP_t- \I \bQ_t)\otimes(\bP_t- \I \bQ_t)\Big)(\bd{x}-\bQ) \Big)  \\
+ &\ \I k^2 \rho \bd{A} \Big(-2(\bP\cdot \bQ_t)\big((\bP_t- \I \bQ_t)\cdot (\bd{x}-\bQ)\big) + (\bP\cdot \bQ_t)^2 \Big)  \\
+ &\ (\lambda + 2\mu)\Big(-  \I k \bd{A} + k^2\big((\bd{A}\cdot \bP)(\bd{x}-\bQ)  \\
 &\ +(\bd{A}\otimes \bP)(\bd{x}-\bQ) + \I\big((\bd{x}-\bQ)\otimes(\bd{x}-\bQ)\big)\bd{A} \Big) \\
-&\ 2\I \mu k \bd{A} - \I\mu k^2\Big(((\bd{x}-\bQ)\cdot \bd{A})(\bd{x}-\bQ) - |\bd{x}-\bQ|^2\bd{A}\Big) \\
-&\  \mu k^2\Big((\bP\cdot \bd{A})(\bd{x}-\bQ) - (\bP\cdot(\bd{x}-\bQ))\bd{A}\Big),
\end{aligned}
\end{equation}
where $\otimes$ means the tensor product, e.g., $(\bd{A}\otimes \bd{P})_{jl}=A_jP_l$.

\noindent Expanding $\rho(\bd{x})$ around $\bQ$ and truncating at order third order,
\begin{align}
\rho(\bd{x}) =& \ \rho + \partial_{\bQ}\rho \cdot(\bd{x}-\bQ)+ \ds\frac{1}{2}(\bd{x}-\bQ)\cdot \partial_{\bQ\bQ}^2\rho(\bd{x}-\bQ),
\end{align}
and noticing that $\rho(\bd{Q}) c^2(\bd{Q})=\lambda+2\mu$ is constant, one has
\begin{align}\label{density:1}
\partial_{\bQ}\rho c^2 + 2c \partial_{\bQ}c\rho = 0,\quad \partial_{\bQ}\rho = -\ds\frac{2c \partial_{\bQ}c\rho}{c^2}.
\end{align}
Taking the second derivative for $\partial_{\bQ\bQ}\rho$ and substituting eq.~\eqref{density:1} bring
\begin{equation}\label{eq:rho_taylor}
\rho(\bd{x}) =p  \rho -\ds\frac{2\partial_{\bQ}c\rho}{c}\cdot(\bd{x}-\bQ)  + (\bd{x}-\bQ)\cdot\left(3\ds\frac{\partial_{\bQ}c\partial_{\bQ}c^T\rho}{c^2}  - \ds\frac{\partial_{\bQ\bQ}c\rho}{c}\right)(\bd{x}-\bQ).
\end{equation}
Plugging eqs,~\eqref{eq:char_plus} and \eqref{eq:rho_taylor} into eq.~\eqref{FGA:1}, and dividing by $\rho$ yield, in componentwise form,
\begin{equation}\label{FGA:4}
\begin{aligned}
2 k \partial_tA_i c|\bP| \sim & k A_i \Big(c (\partial_{\bQ}c)_j P_j +\I c^2\Big) - \ds \I c^2 k A_i  \\
&\ +k^2 A_i(x-Q)_jM_{jl}(x-Q)_l  - 2k^2A_ic^2P_j (x-Q)_j + \ds c^2N_{i} \\
 &\ - \ds\frac{2\I k\mu}{\rho}A_i - \ds\frac{k^2\mu}{\rho}\Big(P_jA_j(x-Q)_i - P_j(x-Q)_j A_i\Big)\\
&\ -\ds\frac{\I k^2\mu}{\rho}\Big((x-Q)_jA_j(x-Q)_i - |\bd{x}-\bQ|^2A_i\Big),
\end{aligned}
\end{equation}
where $M_{jl}$ and $N_{i}$ are given as follows,
\begin{equation}\label{fga_fact:1}
\begin{aligned}
M_{jl} =&\  -\I|\bP|^2c(\partial^2_{\bQ}c)_{jk} + c\big(P_l +P_j\big)(\partial_{\bQ}c)_j-\I c^2\ds\frac{P_jP_l}{|\bP|^2},  \\
N_{i} = &\ A_j P_j(x-Q)_i + A_iP_j(x-Q)_j +\I(x-Q)_i(x-Q)_jA_j.
\end{aligned}
\end{equation}

\noindent Assuming that $\bd{A} = a_{\tp}\bd{\hat{N}}_{\tp}$, with $\displaystyle \bd{\hat{N}}_\tp=\frac{\bd{P}}{\abs{\bd{P}}}$, then \begin{equation}
\partial_t \bd{A} = \partial_ta_\tp\bd{\hat{N}}_{\tp} + a_{\tp}\partial_t\bd{\hat{N}}_{\tp}=\partial_ta_\tp \frac{\bd{P}}{\abs{\bd{P}}} + a_{\tp}\partial_t\Bigl(\frac{\bd{P}}{\abs{\bd{P}}}\Bigr).
\end{equation}
Plugging this into eq.~\eqref{FGA:4} with the ray equations \eqref{eq:char_plus}, using the P-wave velocity eq.~\eqref{eq:ps_speed} and grouping in powers of $(\bd{x}-\bQ)$ produce
\begin{align*}
2k\partial_t a_{\tp}c\abs{\bd{P}}P_i \sim &\ -2ka_{\tp}c|\bP|^2 \partial_t \Bigl(\frac{P_i}{\abs{\bd{P}}}\Bigr) + k a_{\tp} c (\partial_{\bQ}c)_jP_j P_i \\
&\  -\ds\frac{2\I k\mu P_i }{\rho}a_{\tp} + k^2a_{\tp}|\bP|^2\left(c^2 - \ds\frac{\mu}{\rho} \right)(x-Q)_i \\
 &\ + k^2 a_{\tp} \left(\ds\frac{\mu}{\rho}-c^2 \right)P_jP_i(x-Q)_j  	+ k^2\ds\frac{\I\mu a_{\tp}}{\rho}|x-Q|^2P_i \\
 &\ + k^2\I a_{\tp}P_j\left(c^2 - \ds\frac{\mu}{\rho}\right)(x-Q)_j(x-Q)_i + k^2a_{\tp}P_i(x-Q)_j M_{jl}(x-Q)_l.
\end{align*}
Denoting $\partial_{\bd{z}}=(\partial_1,\partial_2,\partial_3)$ for an ease of notations, applying eq.~\eqref{lemma:1} and dropping the lower order terms yield
\begin{equation}\label{eq:1}
\begin{aligned}
2\partial_ta_{\tp}c\abs{\bd{P}}P_i \sim &\ -2ka_{\tp}c|\bP|^2 \partial_t \Bigl(\frac{P_i}{\abs{\bd{P}}}\Bigr) + a_{\tp}c (\partial_{\bQ}c)_jP_j P_i - \ds\frac{2\I\mu a_{\tp}}{\rho}P_i \\
- &\ 
\partial_{l}\left(\Big(|\bP|^2 - P_jP_i\Big)a_{\tp}\left(\ds\frac{\mu}{\rho} - c^2\right) Z^{-1}_{jl}\right)\\
+ &\ 
\I a_{\tp}P_j\left(c^2 - \ds\frac{\mu}{\rho}\right) Z_{il}^{-1}\partial_{l}Q_j
+ \ds\frac{\I\mu a_{\tp}}{\rho}P_i Z_{jl}^{-1}\partial_{l}Q_j \\
+ &\ 
a_{\tp}P_i\left(c(\partial_{\bQ}c)_jP_r  +cP_j(\partial_{\bQ}c)_j -\I c^2\ds\frac{P_jP_r}{|\bP|^2}\right)cZ_{rl}^{-1}\partial_{l}Q_j \\
 - &\ a_{\tp}\I P_i|\bP|^2c(\partial^2_{\bQ}c)_{jr} Z_{rl}^{-1}\partial_{l}Q_j.
\end{aligned}
\end{equation}
To derive an ordinary differential equation (ODE) instead of a partial differential equation (PDE) for $a_{\tp}$, one needs to simplify the terms containing $\partial_ka_{\tp}$ as
\begin{equation}\label{eq:a_der}
\Big(|\bP|^2 - P_jP_i\Big)\partial_l\left(a_{\tp}\left(\ds\frac{\mu}{\rho} - c^2\right)
 Z^{-1}_{jl}\right).
\end{equation}
Recall that eq.~\eqref{eq:1} holds in the sense of integral form \eqref{eq:sim}, and now we shall consider a strong form of eq.~\eqref{eq:1}, i.e., equate the integrands of the integrals on both sides. After taking the dot product of integrands with $\bP$, one has
\begin{align*}
 2\ds\frac{\partial_t a_{\tp}}{a_{\tp}}
= &\ 
\ds2\frac{(\partial_{\bQ}c)_jP_j}{|\bP|} - \ds\frac{2\mu\I}{\rho c|\bP|} + \ds\frac{\I\mu}{\rho c|\bP|}\left(\delta_{ij} - \ds\frac{P_jP_i}{|\bP|^2}\right) Z_{il}^{-1}\partial_{l}Q_j \\
+ &\ 
\ds\frac{1}{c}\left(c^2 - \ds\frac{\mu}{\rho}\right)\ds\frac{P_i}{|\bP|}\partial_{l}\left(\ds\frac{P_jP_i}{|\bP|^2}\right) Z_{jl}^{-1} + \ds\frac{1}{|\bP|}(\partial_{\bQ}c)_jP_k Z_{kl}^{-1}\partial_{l}Q_j  \\
+&\ \ds\frac{1}{|\bP|}P_j(\partial_{\bQ}c)_j Z_{kl}^{-1}\partial_{l}Q_j -\I|\bP|(\partial^2_{\bQ}c)_{jk} Z_{kl}^{-1}\partial_{l}Q_j,
\end{align*}
where the terms \eqref{eq:a_der} actually become zero since $\bP\cdot(|\bP|^2 - \bP\otimes\bP)=0$, and we have used the fact that
\begin{equation*}
\frac{P_i}{\abs{\bd{P}}}\partial_t \Bigl(\frac{P_i}{\abs{\bd{P}}}\Bigr)=\frac{1}{2}\biggl( \frac{P_i}{\abs{\bd{P}}}\partial_t \Bigl(\frac{P_i}{\abs{\bd{P}}}\Bigr) + \partial_t \Bigl(\frac{P_i}{\abs{\bd{P}}}\Bigr) \frac{P_i}{\abs{\bd{P}}}\biggr)=\partial_t\biggl(\frac{P_i}{\abs{\bd{P}}}\frac{P_i}{\abs{\bd{P}}}\biggr)=0,
\end{equation*}
which implies $\displaystyle \partial_t \Bigl(\frac{P_i}{\abs{\bd{P}}}\Bigr)P_i=0$.

\noindent Since $ Z = \partial_{\bz}(\bQ + \I \bP)$ by eq.~\eqref{eq:op_zZ_app}, $\partial_t  Z=\partial_t\partial_{\bz}\bQ + \I\partial_t\partial_{\bz}\bP$. Then eq.~\eqref{eq:char_plus} implies
\begin{equation}\label{eq:DzDt}
\begin{aligned}
\partial_t\partial_{\bz}\bQ
=&\ \partial_{\bz}\bQ \ds\frac{\partial_{\bQ}c\otimes \bP}{|\bP|} + c\partial_{\bz}\bP \left(\frac{\Id_3}{|\bP|} - \ds\frac{\bP \otimes \bP}{|\bP|^3}\right), \\
\partial_t\partial_{\bz}\bP
=&\ -|\bP|\partial_{\bz}\bQ \partial^2_{\bQ}c - \partial_{\bz}\bP\ds\frac{\bP\otimes \partial_{\bQ} c}{|\bP|}.
\end{aligned}
\end{equation}
Using eq.~\eqref{eq:DzDt} for further simplifications give
\begin{equation}
\begin{aligned}
2\ds\frac{\partial_t a_{\tp}}{a_{\tp}}
= &\ %
 \ds 2\frac{(\partial_{\bQ} c)_iP_i}{|\bP|} + \tr\left( Z^{-1} \partial_t Z\right) \\
 &\ + %
\ds\frac{1}{c}\left(c^2 - \ds\frac{\mu}{\rho}\right)\ds\frac{P_i}{|\bP|}\partial_{l}\left(\ds\frac{P_jP_i}{|\bP|^2}\right) Z_{jl}^{-1} - \ds\frac{1}{ c|\bP|}\left(c^2 - \ds\frac{\mu}{\rho} \right)\left(\delta_{ij} - \ds\frac{P_jP_i}{|\bP|^2}\right) Z_{il}^{-1}\partial_{l}P_j,
\end{aligned}
\end{equation}
where the last two terms can be grouped as
\begin{align}
\ds\frac{1}{c|\bP|}\left(c^2 - \ds\frac{\mu}{\rho}\right)\partial_{l}\left(P_i\Big(\ds\frac{P_jP_i}{|\bP|^2}-\delta_{ij}\Big)\right) Z_{jl}^{-1} =0,
\end{align}
which implies a clean ODE for $a_{\tp}$ as in eq.~\eqref{eq:amp_P},
\begin{equation}
\frac{\ud a_{\tp}}{\ud t} = a_\tp\left(\ds\frac{\partial_{\bQ}c\cdot \bP}{|\bP|} + \ds\frac{1}{2}\tr\left( Z^{-1} \frac{\ud Z}{\ud t}\right)\right).
\end{equation}

Similarly, one can derive the prefactor equations for SV- and SH-waves by assuming $\bd{A}=a_{\tsv}\bd{\hat{N}}_{\tsv}+a_{\tsh}\bd{\hat{N}}_{\tsh}$ with $\bd{\hat{N}}_\tsv\perp \bd{P}$, $\bd{\hat{N}}_\tsh\perp \bd{P}$ and $\bd{\hat{N}}_\tsv\perp \bd{\hat{N}}_\tsh$ in eq.~\eqref{solform:1}. The calculations will be essentially the same as the prefactor equation for P-waves except that one will have the diabatic coupling terms of $\bd{\hat{N}}_{\tsv}$ and $\bd{\hat{N}}_{\tsh}$ as shown below,
\begin{equation}\label{eq:2}
\begin{aligned}
& \frac{\ud a_{\tsv}}{\ud t} = a_{\tsv}\biggl(\frac{\partial_{\bd{Q}_{\ts}}c_{\ts}\cdot \bd{P}_{\ts}}{|\bd{P}_{\ts}|} + \frac{1}{2}\tr\Bigl(Z_{\ts}^{-1}\frac{\ud Z_{\ts}}{\ud t}\Bigr)\biggr)-a_{\tsh}\biggl(\frac{\ud \bd{\hat{N}}_{\tsh}}{\ud t}\cdot\bd{\hat{N}}_\tsv+m_{\tsh\rightarrow\tsv}\biggr), \\
& \frac{\ud a_{\tsh}}{\ud t} = a_{\tsh}\biggl(\frac{\partial_{\bd{Q}_{\ts}}c_{\ts}\cdot \bd{P}_{\ts}}{|\bd{P}_{\ts}|} + \frac{1}{2}\tr\Bigl(Z_{\ts}^{-1}\frac{\ud Z_{\ts}}{\ud t}\Bigr)\biggr)-a_{\tsv}\biggl(\frac{\ud \bd{\hat{N}}_{\tsv}}{\ud t}\cdot\bd{\hat{N}}_\tsh+m_{\tsv\rightarrow\tsh}\biggr),
\end{aligned}
\end{equation}
where the interaction terms are given by
\begin{equation*}
\begin{aligned}
&m_{\tsh\rightarrow\tsv} = \I \frac{\lambda+\mu}{\rho c_\ts\abs{P_{\ts}}}\Bigl(\bd{\hat{N}}_{\tsv}\cdot(Z_\ts^{-1}\partial_{\bd{z}}\bd{Q}_s)\bd{\hat{N}}_{\tsh} - \bd{\hat{N}}_{\tsh}\cdot(Z_\ts^{-1}\partial_{\bd{z}}\bd{Q}_s)\bd{\hat{N}}_{\tsv} \Bigr), \\ & m_{\tsv\rightarrow\tsh}=-m_{\tsh\rightarrow\tsv}.
\end{aligned}
\end{equation*}
Also, note that by $\bd{\hat{N}}_\tsv\perp \bd{\hat{N}}_\tsh$, one has that $\displaystyle \frac{\ud \bd{\hat{N}}_{\tsh}}{\ud t}\cdot\bd{\hat{N}}_\tsv + \frac{\ud \bd{\hat{N}}_{\tsv}}{\ud t}\cdot\bd{\hat{N}}_\tsh =0$.

\noindent Next, we shall show that $m_{\tsh\rightarrow\tsv}=m_{\tsv\rightarrow\tsh}=0$ by proving that $Z_\ts^{-1}\partial_{\bd{z}}\bd{Q}_s$ is symmetric using the following argument. Eq.~\eqref{eq:symplectic} implies, with the subscript $\ts$ omitted for convenience,
\begin{align}
& \partial_{\bd{q}}\bd{Q}(\partial_{\bd{q}}\bd{P})^\TT - \partial_{\bd{q}}\bd{P}(\partial_{\bd{q}}\bd{Q})^\TT = 0_{3\times 3}, \label{eq:p1}\\
& \partial_{\bd{q}}\bd{Q}(\partial_{\bd{p}}\bd{P})^\TT - \partial_{\bd{q}}\bd{P}(\partial_{\bd{p}}\bd{Q})^\TT = \Id_3, \label{eq:p2} \\
& \partial_{\bd{p}}\bd{Q}(\partial_{\bd{q}}\bd{P})^\TT - \partial_{\bd{p}}\bd{P}(\partial_{\bd{q}}\bd{Q})^\TT = - \Id_3, \label{eq:p3} \\
& \partial_{\bd{p}}\bd{Q}(\partial_{\bd{p}}\bd{P})^\TT - \partial_{\bd{p}}\bd{P}(\partial_{\bd{p}}\bd{Q})^\TT = 0_{3\times 3}, \label{eq:p4}
\end{align}
where $0_{3\times 3}$ is $3$-by-$3$ zero matrix.

\noindent Eq.~\eqref{eq:p1}$-\I\times$ Eq.~\eqref{eq:p3} gives
\begin{equation}\label{eq:p5}
\partial_{\bd{z}}\bd{Q}(\partial_{\bd{q}}\bd{P})^\TT - \partial_{\bd{z}}\bd{P}(\partial_{\bd{q}}\bd{Q})^\TT = \I\Id_3.
\end{equation}

\noindent Eq.~\eqref{eq:p2}$-\I\times$ Eq.~\eqref{eq:p4} gives
\begin{equation}\label{eq:p6}
\partial_{\bd{z}}\bd{Q}(\partial_{\bd{p}}\bd{P})^\TT - \partial_{\bd{z}}\bd{P}(\partial_{\bd{p}}\bd{Q})^\TT = \Id_3.
\end{equation}

\noindent Eq.~\eqref{eq:p5}$-\I\times$ Eq.~\eqref{eq:p6} gives
\begin{equation*}
\partial_{\bd{z}}\bd{Q}(\partial_{\bd{z}}\bd{P})^\TT - \partial_{\bd{z}}\bd{P}(\partial_{\bd{z}}\bd{Q})^\TT = 0_{3\times 3}.
\end{equation*}
Combined with $\partial_{\bd{z}}\bd{Q}(\partial_{\bd{z}}\bd{Q})^\TT - \partial_{\bd{z}}\bd{Q}(\partial_{\bd{z}}\bd{Q})^\TT = 0_{3\times 3}$, one has
\begin{equation*}
\partial_{\bd{z}}QZ^\TT-Z(\partial_{\bd{z}}Q)^\TT=0,
\end{equation*}
which implies $Z^{-1}\partial_{\bd{z}}Q=(\partial_{\bd{z}}Q)^\TT(Z^\TT)^{-1}=(\partial_{\bd{z}}Q)^\TT(Z^{-1})^\TT=(Z^{-1}\partial_{\bd{z}}Q)^\TT$. Therefore, $Z^{-1}\partial_{\bd{z}}Q$ is symmetric, and then $m_{\tsh\rightarrow\tsv}=m_{\tsv\rightarrow\tsh}=0$, which brings eqs.~\eqref{eq:amp_SV} and \eqref{eq:amp_SH} by eq.~\eqref{eq:2}.

\section{Derivation of the interface conditions for $\partial_{\bd{z}}\bd{Q}$ and $\partial_{\bd{z}}\bd{P}$}\label{app:interface}
The derivation of the interface conditions for $\partial_{\bd{z}}\bd{Q}_{\tp,\ts}$ and $\partial_{\bd{z}}\bd{P}_{\tp,\ts}$ requires to use the conservation of level set functions designed in the Eulerian frozen Gaussian approximation formula \citep{LuYa:MMS,WeYa:12}, where the idea is to use the following Liouville operator to describe the dyanmics of Gaussian wave packet on phase plane,
\begin{equation}\label{eq:Liou}
\cL_{\tp,\ts} = \partial_t + \partial_{\bd{p}} H_{\tp,\ts}\cdot\partial_{\bd{q}} - \partial_{\bd{q}} H_{\tp,\ts}\cdot\partial_{\bd{p}}
\end{equation}
whose corresponding characteristic equations are given by the Hamiltonian systems eqs.~\eqref{eq:char_plus} and \eqref{eq:char_minus} with $H_{\tp,\ts}=\pm c_{\tp,\ts}(\bd{q})\abs{\bd{p}}$. For example, the prefactor equation of the P-wave for the ``$+$'' wave propagation direction is given by, in the Eulerian formulation,
\begin{equation*}
\cL a_{\tp} = a_{\tp}\biggl(\frac{\partial_{\bQ_{\tp}}c_{\tp}(\bQ_{\tp})\cdot \bP_{\tp}}{|\bP_{\tp}|} +\frac{1}{2}\tr\left( Z_{\tp}^{-1}\partial_t  Z_{\tp}\right)\biggr).
\end{equation*}

We shall derive the interface conditions of $\partial_{\bd{z}}\bd{Q}$ and $\partial_{\bd{z}}\bd{P}$ for the transmitted P-wave and the ``$+$'' wave propagation direction, and omit the subscript ``$\tp$'' in the following derivation for a sake of simplicity. Consider a level set function $\boldsymbol{\phi}(t,\bq,\bp) = (\phi_{1},\phi_{2},\phi_{3})$ which satisfies
\begin{equation}\label{eq:phi}
\cL \boldsymbol{\phi} = 0, \quad\text{with}\quad \boldsymbol{\phi}(0,\bq,\bp) = \bp + \I\bq,
\end{equation}
then the Eulerian formulation of FGA \citep{LuYa:MMS} shows that
\begin{equation}\label{eq:eulerfor}
\partial_{\bz}\bQ = (\partial_{\bp}\boldsymbol{\phi})^\TT,\indent \partial_{\bz}\bP = -(\partial_{\bq}\boldsymbol{\phi})^\TT.
\end{equation}
We will follow the strategy described in \citet{chai2018tomo,WeYa:12}, and consider the case illustrated in Fig.~\ref{fig:interface}, where the level set functions $\boldsymbol{\phi}^{\RE,\TR}$ for the transmitted P-waves satisfy the same evolution as $\boldsymbol\phi$ in eq.~\eqref{eq:phi} with the following interface conditions
\begin{equation}\label{eq:intercond}
\boldsymbol{\phi}^{\TR}(t,\bq^\TR,\bp^\TR) = \boldsymbol{\phi}^{\TR}(t,\bq^\IN,\bp^\IN).
\end{equation}
Differentiating eq.~\eqref{eq:intercond} by the definition of partial derivatives and chain rule, and making use of eq.~\eqref{eq:p:int} yield
\begin{equation}\label{eq:inter_Dpphi}
\begin{aligned}
\partial_{p_x} \boldsymbol{\phi}^{\TR}(t,\bq^\TR,\bp^\TR) =&\ \left(n_1^{-2} - 1\right)\frac{p_x}{p_z}\partial_{p_z} \boldsymbol{\phi}^{\TR}(t,\bq^\IN,\bp^\IN) + \partial_{p_x} \boldsymbol{\phi}^{\TR}(t,\bq^\IN,\bp^\IN),  \\
\partial_{p_y} \boldsymbol{\phi}^{\TR}(t,\bq^\TR,\bp^\TR) =&\ \left(n_1^{-2} - 1\right)\frac{p_y}{p_z}\partial_{p_z} \boldsymbol{\phi}^{\TR}(t,\bq^\IN,\bp^\IN) + \partial_{p_y} \boldsymbol{\phi}^{\TR}(t,\bq^\IN,\bp^\IN), \\
\partial_{p_z} \boldsymbol{\phi}^{\TR}(t,\bq^\TR,\bp^\TR) =&\ n_1^{-2}\frac{p_z^\TR}{p_z}\partial_{p_z} \boldsymbol{\phi}^{\TR}(t,\bq^\IN,\bp^\IN).
\end{aligned}
\end{equation}
Moreover, differentiating eq.~\eqref{eq:intercond} with respect to $t$ gives
\begin{equation}
\partial_t\boldsymbol{\phi}^{\TR}(t,\bq^\TR,\bp^\TR) = \partial_t\boldsymbol{\phi}^{\TR}(t,\bq^\IN,\bp^\IN),
\end{equation}
which implies by eq.~\eqref{eq:Liou}, at the interface,
\begin{align}
\left[\nabla_{\bp}H \cdot \nabla_{\bq}\boldsymbol{\phi}^{\TR}\right]_{q_z = z_0} = \left[\nabla_{\bq}H\cdot \nabla_{\bp}\boldsymbol{\phi}^{\TR}\right]_{q_z = z_0},
\end{align}
with $[\cdot]$ denoting the jump function. Therefore,
\begin{equation}\label{eq:jump_app}
\begin{aligned}
\partial_{q_x} \boldsymbol{\phi}^{\TR}(t,\bq^\TR,\bp^\TR) =&\ \partial_{q_x} \boldsymbol{\phi}^{t}(t,\bq^\IN,\bp^\IN),  \\
\partial_{q_y} \boldsymbol{\phi}^{\TR}(t,\bq^\TR,\bp^\TR) =&\ \partial_{q_y} \boldsymbol{\phi}^{t}(t,\bq^\IN,\bp^\IN),  \\
\partial_{q_z} \boldsymbol{\phi}^{\TR}(t,\bq^\TR,\bp^\TR) =&\ n_1^{2}\frac{p_z}{p^\TR_z}\partial_{q_z} \boldsymbol{\phi}^{\TR}(t,\bq^\IN,\bp^\IN) + (n_1^{2} - 1)\left(\frac{p_x}{p^\TR_z}\partial_{q_x} + \frac{p_y}{p^\TR_z}\partial_{q_y}\right)\boldsymbol{\phi}^{\TR}(t,\bq^\IN,\bp^\IN) \\
  +&\ \frac{|\bp^\TR|}{c^\vee p^\TR_z} \left[\abs{\bp}\nabla_{\bq} c(\bq)\cdot\nabla_{\bp}\boldsymbol{\phi}^\TR(t,\bq,\bp)\right]_{q_z = z_0}.
\end{aligned}
\end{equation}
Then eqs.~\eqref{eq:eulerfor} and \eqref{eq:inter_Dpphi} imply the interface condition for $\partial_z\bQ$ in eq.~\eqref{eq:Inter_DzQDzP}, while eqs.~\eqref{eq:eulerfor} and \eqref{eq:jump_app} imply the interface condition for $\partial_z\bP$ in eq.~\eqref{eq:Inter_DzQDzP} for the transmitted P-wave. The other cases including the interface conditions for the reflected P-wave, transmitted and reflected S-wave can be derived in an essentially same way.

\clearpage
\newpage
%
\bibliographystyle{gji}
\bibliography{fga_ref,paper,SEG}


\end{document}